\documentclass[fleqn,usenatbib]{mnras}

\usepackage{newtxtext,newtxmath}

\usepackage[T1]{fontenc}

\DeclareRobustCommand{\VAN}[3]{#2}
\let\VANthebibliography\thebibliography
\def\thebibliography{\DeclareRobustCommand{\VAN}[3]{##3}\VANthebibliography}

\usepackage{graphicx}	
\usepackage{amsmath}	
\usepackage{tikz,xcolor,hyperref}

\definecolor{lime}{HTML}{A6CE39}
\DeclareRobustCommand{\orcidicon}{%
    \begin{tikzpicture}
    \draw[lime, fill=lime] (0,0) 
    circle [radius=0.16] 
    node[white] {{\fontfamily{qag}\selectfont \tiny ID}};
    \draw[white, fill=white] (-0.0625,0.095) 
    circle [radius=0.007];
    \end{tikzpicture}
    \hspace{-2mm}
}

\newcommand{\orcidEricGaidos}{\href{https://orcid.org/0000-0002-5258-6846}{\orcidicon}}
\newcommand{\orcidvignesh}{\href{https://orcid.org/0000-0003-2310-9415}{\orcidicon}}
\newcommand{\orcidteru}{\href{https://orcid.org/0000-0003-3618-7535}{\orcidicon}}
\newcommand{\orcidravi}{\href{https://orcid.org/0000-0002-5893-2471}{\orcidicon}}
\newcommand{\orcidtom}{\href{https://orcid.org/0000-0001-7139-2724}{\orcidicon}}
\newcommand{\orcidkatherine}{\href{https://orcid.org/0000-0001-6398-8755}{\orcidicon}}
\newcommand{\orcidklaus}{\href{https://orcid.org/0000-0003-0786-2140}{\orcidicon}}
\newcommand{\orcidharakawa}{\href{https://orcid.org/0000-0002-6197-5544}{\orcidicon}}
\newcommand{\orcidkonishi}{\href{https://orcid.org/0000-0003-0114-0542}{\orcidicon}}
\newcommand{\orcidkotani}{\href{https://orcid.org/0000-0001-6181-3142}{\orcidicon}}
\newcommand{\orcidkudo}{\href{https://orcid.org/0000-0002-9294-1793}{\orcidicon}}
\newcommand{\orcidkuzuhara}{\href{https://orcid.org/0000-0002-4677-9182}{\orcidicon}}
\newcommand{\orcidnishikawa}{\href{https://orcid.org/0000-0001-9326-8134}{\orcidicon}}
\newcommand{\orcidomiya}{\href{https://orcid.org/0000-0002-5051-6027}{\orcidicon}}
\newcommand{\orcidjosh}{\href{https://orcid.org/0000-0001-5347-7062}{\orcidicon}}
\newcommand{\orcidtamura}{\href{https://orcid.org/0000-0002-6510-0681}{\orcidicon}}
\newcommand{\orcidvievard}{\href{https://orcid.org/0000-0003-4018-2569}{\orcidicon}}

\title[low-mass star radius-gap planets]{Absence of extended atmospheres in low-mass star radius-gap planets}

\author[V. Krishnamurthy et al.]{Vigneshwaran Krishnamurthy,$^{1,2,3}$\thanks{E-mail: vigneshwaran.krishnamurthy@mcgill.ca}\orcidvignesh
Teruyuki Hirano,$^{1,2}$\orcidteru
Eric Gaidos,$^{4}$\orcidEricGaidos
Bunei Sato,$^{5}$
Ravi Kopparapu,$^{6}$\orcidravi
\newauthor Thomas Barclay,$^{6,7}$\orcidtom
Katherine Garcia-Sage,$^{8}$\orcidkatherine
Hiroki Harakawa,$^{9}$\orcidharakawa
Klaus Hodapp,$^{10}$\orcidklaus
Shane Jacobson,$^{10}$ 
\newauthor Mihoko Konishi,$^{11}$\orcidkonishi
Takayuki Kotani,$^{1,2,12}$\orcidkotani
Tomoyuki Kudo,$^{9}$\orcidkudo
Takashi Kurokawa,$^{1,13}$
\newauthor Masayuki Kuzuhara,$^{1,2}$\orcidkuzuhara
Eric Lopez,$^{6}$
Jun Nishikawa,$^{2,12,1}$\orcidnishikawa
Masashi Omiya,$^{1,2}$\orcidomiya
Joshua E. Schlieder,$^{6}$\orcidjosh
\newauthor Takuma Serizawa,$^{2,13}$
Motohide Tamura,$^{1,2,14}$\orcidtamura
Akitoshi Ueda,$^{1,2,12}$
and Sebastien Vievard,$^{1,9}$\orcidvievard
\\
$^{1}$Astrobiology Center, 2-21-1 Osawa, Mitaka, Tokyo 181-8588, Japan\\
$^{2}$National Astronomical Observatory of Japan, 2-21-1 Osawa, Mitaka, Tokyo 181-8588, Japan\\
$^{3}$Trottier Space Institute at McGill, McGill University, 3550 University Street, Montreal, QC H3A 2A7, Canada\\
$^{4}$Department of Earth Sciences, University of Hawai`i at M\={a}noa, Honolulu, HI 96822, USA\\
$^{5}$Department of Earth and Planetary Sciences, Tokyo Institute of Technology, 2-12-1 Ookayama, Meguro-ku, Tokyo 152-8551, Japan\\
$^{6}$NASA Goddard Space Flight Center, 8800 Greenbelt Road, Greenbelt, MD 20771, USA\\
$^{7}$University of Maryland, Baltimore County, 1000 Hilltop Circle,
Baltimore, MD 21250, USA\\
$^{8}$NASA GSFC, Heliophysics Science Division Greenbelt, Code 674, MD 20771, USA\\
$^{9}$Subaru Telescope, 650 N. Aohoku Place, Hilo, HI 96720, USA\\
$^{10}$University of Hawaii, Institute for Astronomy, 640 N. Aohoku Place, Hilo, HI 96720, USA \\
$^{11}$Faculty of Science and Technology, Oita University, 700 Dannoharu, Oita 870-1192, Japan\\
$^{12}$Department of Astronomy, School of Science, The Graduate University for Advanced Studies (SOKENDAI), 2-21-1 Osawa, Mitaka, Tokyo 181-8588, Japan\\
$^{13}$Institute of Engineering, Tokyo University of Agriculture and Technology, 2-24-16, Nakacho, Koganei, Tokyo, 184-8588, Japan\\
$^{14}$Department of Astronomy, Graduate School of Science, The University of Tokyo, 7-3-1 Hongo, Bunkyo-ku, Tokyo 113-0033, Japan
}

\date{Accepted for MNRAS}

\pubyear{2022}

\begin{document}
\label{firstpage}
\pagerange{\pageref{firstpage}--\pageref{lastpage}}
\maketitle

\begin{abstract}
\textit{Kepler} showed a paucity of planets with radii of 1.5 -- 2 $\mathrm R_{\oplus}$ around solar mass stars but this radius-gap has not been well studied for low-mass star planets. Energy-driven escape models like photoevaporation and core-powered mass-loss predict opposing transition regimes between rocky and non-rocky planets when compared to models depicting planets forming in gas-poor environments. Here we present transit observations of three super-Earth sized planets in the radius-gap around low-mass stars using high-dispersion InfraRed Doppler (IRD) spectrograph on the Subaru 8.2m telescope. The planets GJ 9827 b and d orbit around a K6V star and TOI-1235 b orbits a M0.5 star. We limit any planet-related absorption in the 1083.3\,nm lines of triplet He\,I by placing an upper-limit on the equivalent width of 14.71\,m\AA, 18.39\,m\AA\, and 1.44\,m\AA\, for GJ 9827 b (99\% confidence),  GJ 9827 d (99\% confidence) and TOI-1235 b (95\% confidence) respectively. Using a Parker wind model, we cap the mass-loss at $>$0.25 $\mathrm M_{\oplus}$\,Gyr$^{-1}$ and $>$0.2 $\mathrm M_{\oplus}$\,Gyr$^{-1}$ for GJ 9827 b and d, respectively (99\% confidence), and $>$0.05 $\mathrm M_{\oplus}$\,Gyr$^{-1}$ for TOI-1235 b (95\% confidence) for a representative wind temperature of 5000\,K. Our observed results for the three planets are more consistent with the predictions from photoevaporation and/or core-powered mass-loss models than the gas-poor formation models. However, more planets in the radius-gap regime around the low-mass stars are needed to robustly predict the atmospheric evolution in planets around low-mass stars.
\end{abstract}




\section{Introduction}
NASA's \textit{Kepler} Space Telescope has discovered multitudes of planets in the radius range 1-4 R$_\oplus$ \citep{2012ApJS..201...15H, 2013ApJS..204...24B, dressing_charbonneau_2015} that orbit their host star in less than 100 days. The mass measurements for these planets through radial velocity (RV) follow-up \citep{2014ApJS..210...20M, 2014ApJ...787...80H} and transit timing variations (TTVs) measurements \citep{2013ApJ...772...74W, 2017AJ....154....5H} reveal that most planets smaller than $\sim$1.6 R$_\oplus$ have rocky compositions with no inflated gaseous envelopes. The existence of a radius-gap (a.k.a radius valley), a low-occupancy region in the occurrence rate distribution for close-in planets (P$<$100 days) at 1.5--2.0 R$_\oplus$ has been predicted and resolved for solar-type stars \citep{fulton2017}. This radius-gap essentially defines a bimodality in the occurrence rate distribution, separating the rocky planets from the non-rocky planets with volatiles. This feature, however has been only partially resolved for planets around low-mass star  \citep[$<$0.75 $\mathrm M_{\odot}$;][]{2018AJ....155..127H, cloutiermenou2020} as the sample-size is relatively small. 

Two distinct classes of models attempt to explain the  origin of this radius-gap: the first class of models, the thermally driven atmospheric escape models (e.g. photoevaporation, core-powered mass-loss) rely on the host star irradiation or planet's internal heat to carve out the radius-gap in the occurrence rate distribution. The X-ray and Extreme UV (XUV) radiation from the host star in the first few hundred million years strip the primordial atmosphere of the close-in planets \citep{2003ApJ...598L.121L, OwenWu2013, 2014ApJ...795...65J, LopezFortney2014, ChenRogers2016, OwenWu2017, 2018ApJ...853..163J, 2018MNRAS.479.5303L, Wu2019}. This photoevaporation process predicts the slope of the radius-gap should vary with orbital period as $r_{p,gap} \propto P^{-0.15}$ \citep{2018MNRAS.479.5303L} while the core-powered thermal atmospheric escape models \citep{2016ApJ...825...29G, 2018MNRAS.476..759G, 2019MNRAS.487...24G, 2020MNRAS.493..792G} predict the slope to be $r_{p,gap} \propto P^{-0.13}$ \citep{2020MNRAS.493..792G}. The second class of models assume that some of the close-in planets are formed intrinsically rocky, i.e. they are born in a gas-poor environment. Dynamical friction in the disk from constant collisions, delays the gas accretion with the core still embedded in protoplanetary disk. This friction continues for a few million years, within which the gaseous disk has almost completely dissipated \citep{2014ApJ...797...95L, 2016ApJ...817...90L, 2018MNRAS.479.5303L}. The radius-gap slope changes sign for these gas-poor formation models and vary with orbital period as $r_{p,gap} \propto P^{0.11}$ \citep{2018MNRAS.479.5303L}.

In the case of low-mass star planets, both these classes of models carve out an unique transition region in the radius-period space that has opposition predictions on the nature of planets i.e. rocky or non-rocky \citep[see Figure 15;][]{cloutiermenou2020}. Planets with measurements of mass and radius within this region can provide robust inferences to distinguish between various proposed physical mechanisms for the formation of the radius-gap. The complete \textit{Kepler} planets still does not provide enough samples in the radius-gap to disentangle the predictions from various models \citep{2020AJ....160..108B}. NASA's \textit{Transiting Exoplanet Survey Satellite}, \textit{TESS} mission is expected to add several thousand planets in this radius regime around low-mass stars \citep[mid-K to mid-M type;][]{2018ApJS..239....2B, 2019AJ....157..113B}. Here we present the spectroscopic transit observations of three planets: TOI-1235 b \citep{toi-1235b_1, toi-1235b_2}, a super-Earth dwelling in the radius-gap around an early M-dwarf \citep[M0.5V;][]{toi-1235b_spec_type1, toi-1235b_spec_type2}, and GJ 9827 b \& GJ 9827 d a couple of super-Earths from a late K-dwarf system \citep[K6V;][]{gj9827_disc1} on either side of radius-gap \citep{gj9827_disc1, gj9827_disc2, gj9827_mass2, gj9827_mass1}. 

Observations of planets in and around the radius-gap are essential in understanding the evolution of planetary atmospheres. The upper atmospheric escape of primordial Hydrogen and Helium atmosphere could be a key mechanism in the transition from non-rocky to rocky planets. And the metastable He\,I in the near infrared (NIR) spectral regime is an excellent marker for probing such upper atmospheres \citep{oklopcichirata}. The He\,I triplet at 1083.3\,nm lines can be observed with ground-based high-resolution spectrographs \citep{2000ApJ...537..916S}. The He\,I triplet lines originate from the $\mathrm 2^{3}S$ level, a metastable state where the Helium atoms are populated by the process of recombination of ionized Helium and depopulated by the collision de-excitation \citep{oklopcichirata}. In the last few years, several planets have been detected with the neutral He\,I in their atmosphere \citep[e.g., ][]{2018Natur.557...68S, 2019A&A...623A..58A, 2020ApJ...894...97N, 2020A&A...641A.161Z, 2021ApJ...909L..10P, zhang2022, 2022AJ....163...68Z}. Several non-detection upper-limits have also been estimated \citep[for example:][]{2018Sci...362.1388N, our_paper_trappist1, 2021AJ....162..222V, 2021RNAAS...5..238G}. 

In the next Chapter, we present the transit observations and data reduction of all three planets. This is followed by the analysis of our He\,I study in Chapter 3. In the Chapter 4, we discuss the interpretation of our results. Finally we conclude our results in Chapter 5.

\section{Observations and Data reduction}

In this section, we present our ground-based transit observations and the data reduction technique. Our first target is GJ 9827, a bright (V-mag = 10.39) late K-dwarf hosting three super-Earths \citep{gj9827_disc1, gj9827_disc2}. The second target is TOI-1235, a weakly active M0.5 star \citep{toi-1235b_spec_type1, toi-1235b_spec_type2}, hosting a keystone super-Earth in the radius-gap \citep{toi-1235b_1, toi-1235b_2}. The planetary and stellar parameters are shown in Table 1.

\begin{table}
\label{table1}
\centering
\caption{Planetary and stellar parameters used in our analysis}
\begin{tabular}{lc}
Parameter & Value \\
\hline
\multicolumn{2}{l}{\it Stellar Parameters: GJ 9827} \\
M$_\ast$ (M$_\odot$) & 0.593 $\pm$ 0.018$^a$\\
R$_\ast$ (R$_\odot$) & 0.579 $\pm$ 0.018$^a$ \\
T$_{\mathrm {eff}}$ & 4294 $\pm$ 52$^a$\\
V$_{\mathrm {sys}}$ (km s$^{-1}$) & 32.1 $^b$\\
Age (Gyr) & 6.05$^{+2.50}_{-2.89}$ $^a$\\
\hline
\multicolumn{2}{l}{\it Planet parameters: GJ 9827 b} \\
R$_\mathrm p$(R$_\oplus$) & 1.529 $\pm$ 0.058$^a$ \\
M$_\mathrm p$(M$_\oplus$) & 4.87 $\pm$ 0.37$^a$ \\
$\mathrm \rho_p$ (g cm$^{-3}$) & 7.47$^{+1.1}_{-0.95}$ $^a$ \\
a (AU) & 0.01866 $\pm$ 0.00019$^a$\\
Period (days) & 1.2089765 $\pm$ 2.3e-06$^a$ \\
T$_{\mathrm C}$ (BJD$_{\mathrm {TBD}}$) &  2457738.82586$^{+0.00026}_{-0.00026}$ $^c$\\
T$_{\mathrm 14}$ (days) &  0.05270$^{+0.00093}_{-0.00083}$ $^c$ \\
K$_{\mathrm P}$ (km s$^{-1}$) & 166.3 $\pm$ 1.693 $^d$\\
K$_{\mathrm s}$ (m s$^{-1}$) & 3.5 $\pm$ 0.32 $^a$\\
e & 0.0 $^a$ \\
i (deg) & 86.07$^{+0.41}_{-0.34}$ $^c$\\
\hline
\multicolumn{2}{l}{\it Planet parameters: GJ 9827 d} \\
R$_\mathrm p$(R$_\oplus$) & 1.955 $\pm$ 0.075$^a$ \\
M$_\mathrm p$(M$_\oplus$) & 3.42 $\pm$ 0.62$^a$ \\
$\mathrm \rho_p$ (g cm$^{-3}$) & 2.51$^{+0.57}_{-0.51}$ $^a$ \\
a (AU) & 0.0555 $^{+0.00055}_{-0.00057}$ $^a$ \\
Period (days) & 6.20183 $\pm$ 1e-05$^a$ \\
T$_{\mathrm C}$ (BJD$_{\mathrm {TBD}}$) &  2457740.96111  $\pm$ 0.00044 $^e$\\
T$_{\mathrm 14}$ (days) &  0.05095$^{+0.00147}_{-0.00122}$ $^c$ \\
K$_{\mathrm P}$ (km s$^{-1}$) & 98.2 $\pm$ 0.973 $^d$\\
K$_{\mathrm s}$ (m s$^{-1}$) & 1.63 $\pm$ 0.31 $^a$\\
e & 0.0 $^a$ \\
i (deg) & 87.443$^{+0.045}_{-0.045}$ $^c$\\
\hline
\hline
\multicolumn{2}{l}{\it Stellar Parameters: TOI-1235} \\
M$_\ast$ (M$_\odot$) & 0.630 $\pm$ 0.024$^f$\\
R$_\ast$ (R$_\odot$) & 0.619 $\pm$ 0.019$^f$ \\
T$_{\mathrm {eff}}$ & 3997 $\pm$ 51$^f$\\
V$_{\mathrm {sys}}$ (km s$^{-1}$) & -27.75 $^g$\\
Age (Gyr) & 0.6-10$^f$\\
\hline
\multicolumn{2}{l}{\it Planet parameters: TOI-1235 b} \\
R$_\mathrm p$(R$_\oplus$) & 1.694$^{+0.080}_{-0.077}$ $^f$\\
M$_\mathrm p$(M$_\oplus$) & 5.90$^{+0.62}_{-0.61}$ $^f$ \\
$\mathrm \rho_p$ (g cm$^{-3}$) & 6.7$^{+1.3}_{-1.1}$ $^f$ \\
a (AU) & 0.03826 $^{+0.00048}_{-0.00049}$ $^f$ \\
Period (days) & 3.444717$^{+0.000040}_{-0.000042}$ $^f$ \\
T$_{\mathrm C}$ (BJD$_{\mathrm {TBD}}$) &  2458845.51696$^{+0.00099}_{-0.00098}$ $^h$\\
T$_{\mathrm 14}$ (hours) &  1.84$^{+0.09}_{-0.16}$ $^h$ \\
K$_{\mathrm P}$ (km s$^{-1}$) & 120.83 $\pm$ 1.516 $^i$\\
K$_{\mathrm s}$ (m s$^{-1}$) & 4.11 $^{+0.43}_{-0.50}$ $^h$\\
e & $<$0.15 $^h$ \\
i (deg) & 88.1$^{+0.8}_{-0.9}$ $^h$\\
\hline
\noalign{\smallskip}
$^a$\cite{gj9827_plan_para} & \multicolumn{1}{l}{$^e$\cite{gj9827_disc2}}\\
$^b$\cite{rvgj9827} & \multicolumn{1}{l}{$^f$\cite{toi-1235b_2}}\\
$^c$\cite{gj9827_mass1} & \multicolumn{1}{l}{$^g$\cite{gaiadr2}} \\
$^d$\cite{gj9827_helium} & \multicolumn{1}{l}{$^h$\cite{toi-1235b_1}} \\
\multicolumn{1}{l}{$^i$ This work}
\end{tabular}
\end{table}

\subsection{Observations}
\subsubsection{GJ 9827 system}
The non-overlapping transits of GJ 9827 b and d were captured on the same night of HST 2020 September 05. We used the InfraRed Doppler (IRD) instrument \citep{2012SPIE.8446E..1TT, 2018SPIE10702E..11K} installed on the 8.2\,m Subaru Telescope, Maunakea, Hawaii for our observation. The fiber-fed high-dispersion IRD spectrograph operates in the wavelength range of 0.95–1.75\,$\mu$m with a spectral resolution of R$\approx$70,000. The observation was carried out as part of an open-use observation program S20B-069 (PI: Krishnamurthy). The airmass during the transit of GJ 9827 d was high (1.5--2.2). Also the seeing became poor during the second half of the night (see Table 2). Integration time was initially set to 180 seconds at the beginning from UT 2020 September 06 08:35. Due to the poor seeing and low elevation, the integration time was increased to 300 seconds and further to 360 seconds, by the end of the night to maintain the same SNR. In between the transits of GJ 9827 b and d, about $\sim$2 hours was spared to observe another target \citep[see][]{our_paper_trappist1}. We also lost two spectra due to technical error, leaving us with a total of sixty six usable frames of GJ 9827 by the end of observation (see Table 2). We also obtained six frames for an A1 rapid rotator, HIP 8296 \citep{1988mcts.book.....H} for telluric corrections by the end of the night.

\subsubsection{TOI-1235 b}

We observed a transit of TOI-1235 b on UT 2021 April 09 under the program S21A-100N (PI: Krishnamurthy) using the same IRD instrument installed on the Subaru telescope. We had nominal seeing on the transit night and we set the integration time to 240 seconds. Due to low elevation, we observed TOI-1235 for only 20 minutes after the expected transit egress. In total, we obtained thirty six spectra of TOI-1235. We then shifted to a A1-type rapid-rotator HIP 60044 \citep{2006AstL...32..759G} for telluric correction. Few more frames of TOI-1235 were obtained on UT 2021 April 17 (eight frames) and on UT 2021 April 19 (eight frames), when there were no transits of TOI-1235 b to disentangle the telluric features from the stellar spectra. The conditions on the various nights of observations are shown in Table 2.

\begin{table*} \label{tab2}
\centering
\caption{Observation summary for our targets.}
\begin{tabular}{lccccc}
\hline
Observation date (UT) & Number of observations & Exposure time (seconds) & Airmass & seeing (arcseconds) & SNR (at 1$\mu$m) \\
\hline
\multicolumn{2}{l}{\it GJ 9827} \\
2020 September 06 & 66 & 300-300-360* & 1.2-1.08-2.2* & 0.8-1.2-1.6* & 125-90-60* \\
\hline
\multicolumn{2}{l}{\it TOI-1235} \\
2021 April 09 & 36 & 240-240-240* & 2-2.5-3* & 0.8-1.0-0.9* & 60-60-65* \\
2021 April 17 & 8 & 240-240-240* & 1.6-1.65-1.7* & 0.51-0.71-0.89* & 60-60-65* \\
2021 April 19 & 8 & 240-240-240* & 1.5-1.5-1.5* & 0.4-0.62-0.65* & 60-60-65* \\
\hline
\multicolumn{2}{l}{* values are denoted for beginning - middle - end of the observation.} \\
\end{tabular}
\end{table*}

\subsection{Data reduction}

The obtained raw spectra from both targets were reduced using the {\tt IRAF} software, following the procedures described in \cite{2020PASJ...72...93H}. This is followed by the precise wavelength calibration with our custom codes using the simultaneously obtained Laser Frequency Comb (LFC) spectra. The reduced 1-d spectra has an SNR-per-pixel of about 110 for GJ 9827 and 60-65 for TOI-1235 at 1\,$\mu$m.

\subsubsection{Telluric correction and transmission spectra}
\label{new_tell_sec}

\begin{figure}
\centering
\includegraphics[width=8.65cm]{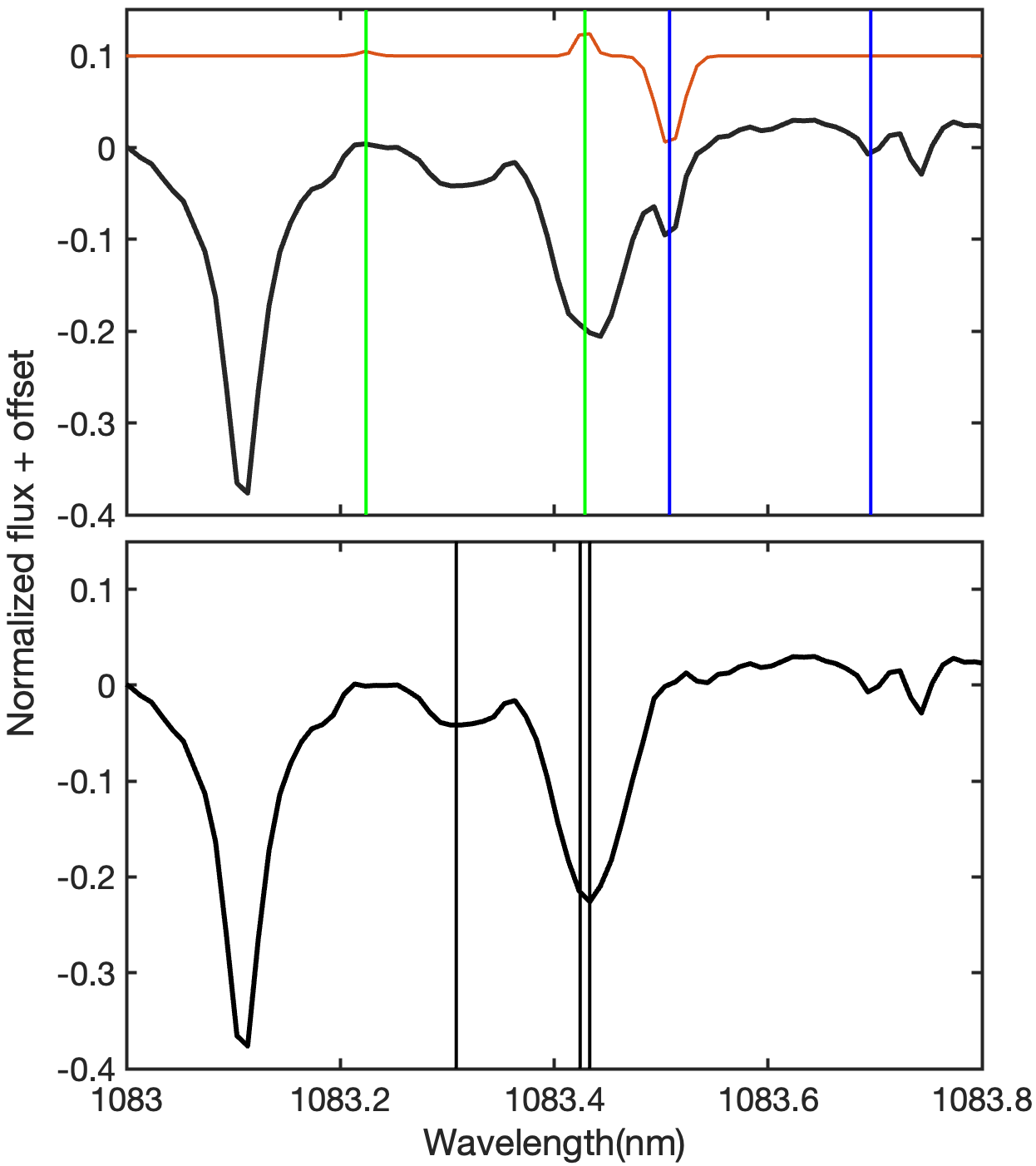}
\caption{Master out-of-transit spectrum (black) of GJ 9827 in observer's rest frame. Top panel shows the constructed telluric template (red) with the line positions of telluric OH (green) and water lines (blue). The bottom panel shows the telluric subtracted master out-of-transit spectrum. The metastable He\,I line positions are marked by black vertical lines.}
\label{fig_new1}
\end{figure}

Before the atomic species search in the atmosphere of the planet, we have to remove the telluric lines. For the GJ 9827 spectra, we modeled the telluric lines from the observed line positions by following the procedures from \cite{2019A&A...623A..58A} and briefly from \cite{2015A&A...577A..62W}. Since there is no additional fiber in IRD to monitor the sky, we used the line positions from literature \citep{telluric_h2o, 2012A&A...543A..92N, telluric_oh} and model the OH emission line from the rapid-rotator observation. The OH emission lines at 1083.22409\,nm and 1083.42872\,nm (in observer rest frame) were modelled using Gaussian profile with best-fit parameters of depth=0.49\%; HWHM=0.014\,nm and depth=2.27\%; HWHM=0.026\,nm respectively. The depth of these OH lines were linearly scaled from other telluric lines on the rapid-rotator spectra that were not overlapping any stellar lines in target spectra. And the water line at 1083.50778\,nm (in observer rest frame) was constructed using a Voigt profile with best-fit parameters as depth=9.38\%; HWHM=0.036\,nm; k$_G$=0.93. The telluric water line at 1083.69591\,nm was so far away from the He\,I lines that it cannot affect the He\,I lines. Hence we did not remove them. The constructed telluric template for GJ 9827 is shown in Figure \ref{fig_new1} (in red).

\begin{figure*}
\centering
\includegraphics[width=17.8cm]{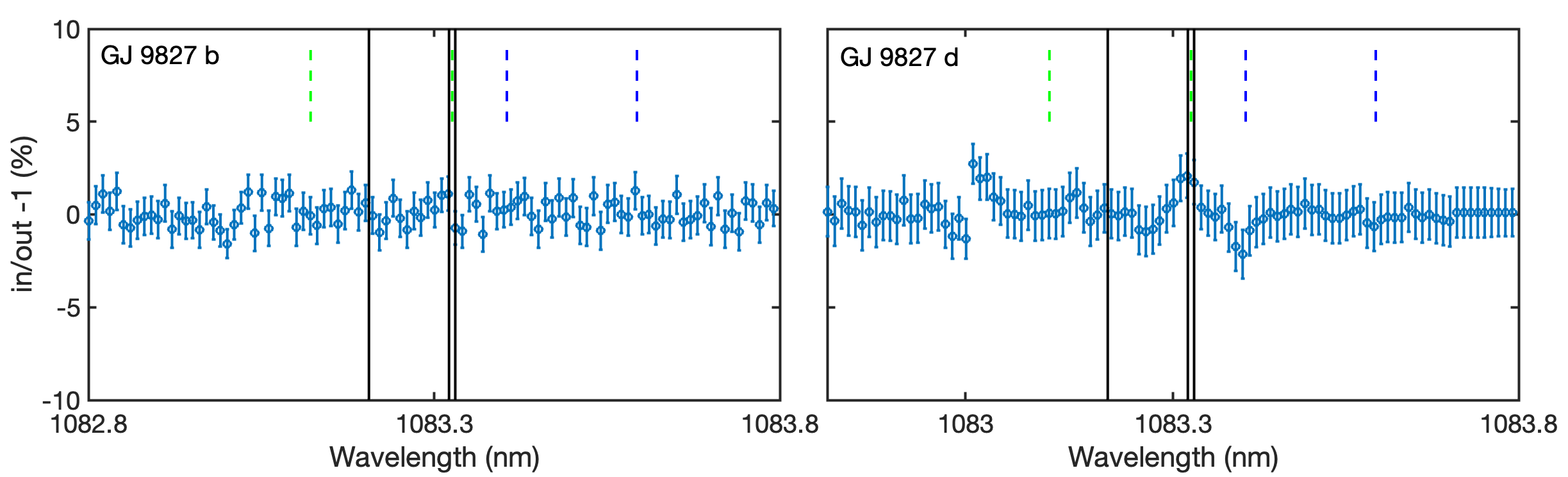}
\caption{The transmission spectra for GJ 9827 b and d in planet's rest frame. The He\,I  line positions are shown in black. The telluric line positions are shown in vertical green (OH lines) and blue ($\mathrm H_2$O) dashed lines.}
\label{fig_new2}
\end{figure*}

The ephemerides of GJ 9827 b and GJ 9827 d were taken from \cite{gj9827_mass1} and \cite{gj9827_disc2} respectively and the planet's updated physical parameters were taken from \cite{gj9827_plan_para} (see Table 1). We removed the last five frames of GJ 9827 from our analysis as the target was at a very low elevation yielding a very low SNR (see Table 2). Then all the out-of-transit spectra and in-transit spectra are separately normalized using the blueward and redward region (1082.6-1082.8\,nm and 1083.8-1084.4\,nm) away from the He\,I lines. The out-of-transit spectra are median-combined to create the master out-of-transit spectra. The telluric template was then subtracted from the median-combined master out-of-transit spectrum to create a telluric-free master stellar spectrum (see bottom panel of Figure \ref{fig_new1}). This was then transferred to the stellar rest frame, accounting for the GJ 9827's systemic RV, stellar reflex motion and Earth's barycentric velocity. For the transit spectra of GJ 9827 b and d, we subtracted the telluric template for each exposure. We then divided each in-transit by the master out-of-transit spectrum to create flux ratio spectra (i.e., transmission spectra). However, the instantaneous change in the water vapor column causes variation in the telluric line strengths. This is especially pronounced in the OH emission lines \citep{kawauchi2018, 2019A&A...623A..58A}. To remove this, we implemented a second telluric correction briefly following the method in Section 2.4 of \cite{2015A&A...577A..62W}. We performed a correlation between telluric template and the individual flux ratio spectra. Since the telluric template only have telluric lines, they would correlate only with the telluric features still present in the individual flux ratio spectra of planet GJ 9827 b and d. To remove the correlations, we linearly fit all the correlated peaks and the telluric template. Each flux ratio was then divided by this fit solution to create telluric-free flux ratios. After this process, we transferred the individual ratios to their respective planetary rest frames by accounting for the planetary RVs and then summed up the ratios to create the flux ratio spectra for GJ 9827 b and d. The flux ratios in percentage is shown in the Figure \ref{fig_new2}.

\begin{figure}
\centering
\includegraphics[width=8.65cm]{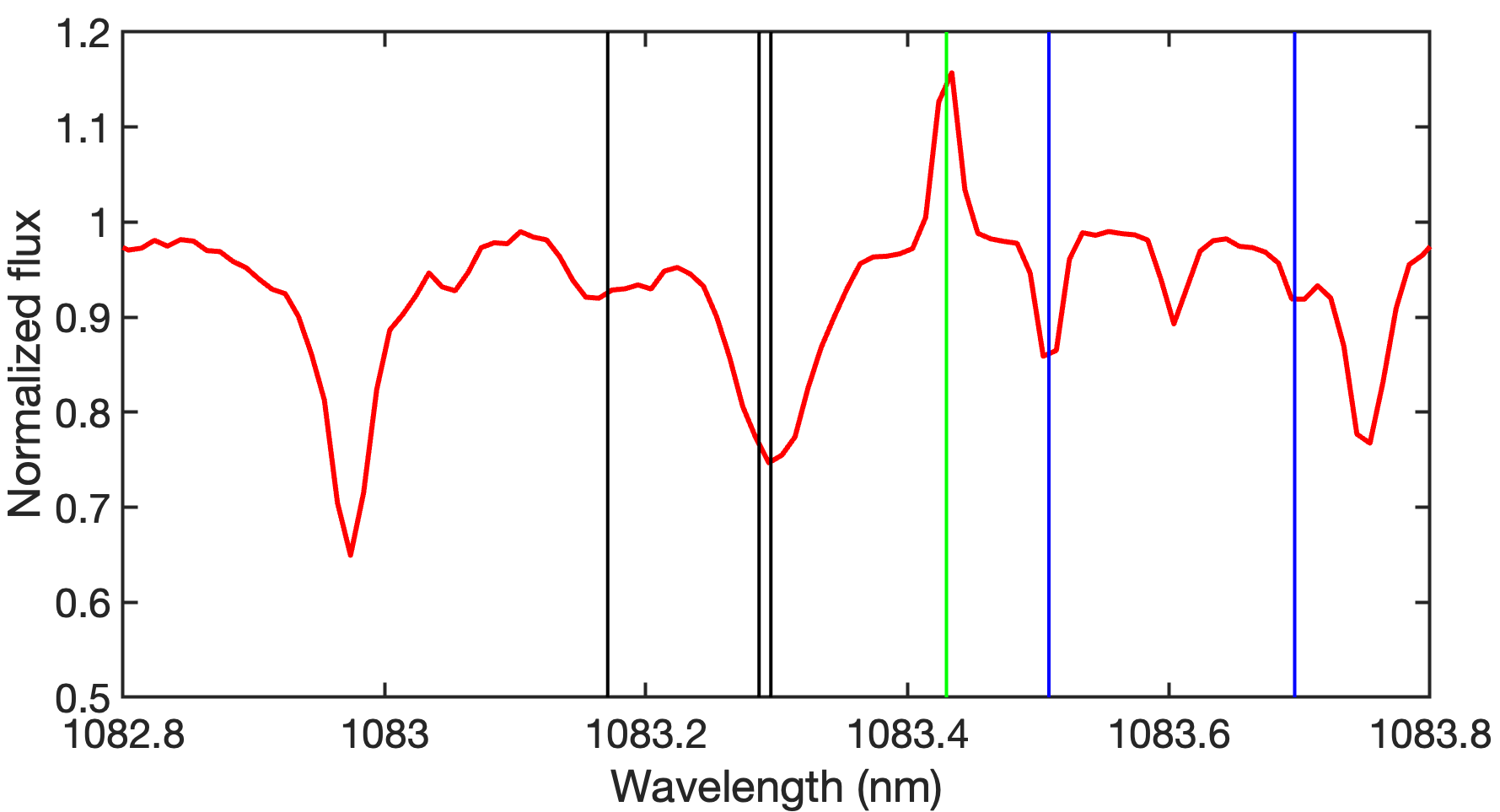}
\caption{Master out-of-transit spectrum of TOI-1235 b in observer's rest frame. The black vertical lines mark the He I triplet positions. The telluric positions of OH emission lines (green) and $\mathrm H_2$O absorption lines (blue) are taken from the literature \citep{telluric_h2o, 2012A&A...543A..92N, telluric_oh}. Here the telluric lines are far away from the He\,I lines.}
\label{fig4}
\end{figure}

\begin{figure}
\centering
\includegraphics[width=8.65cm]{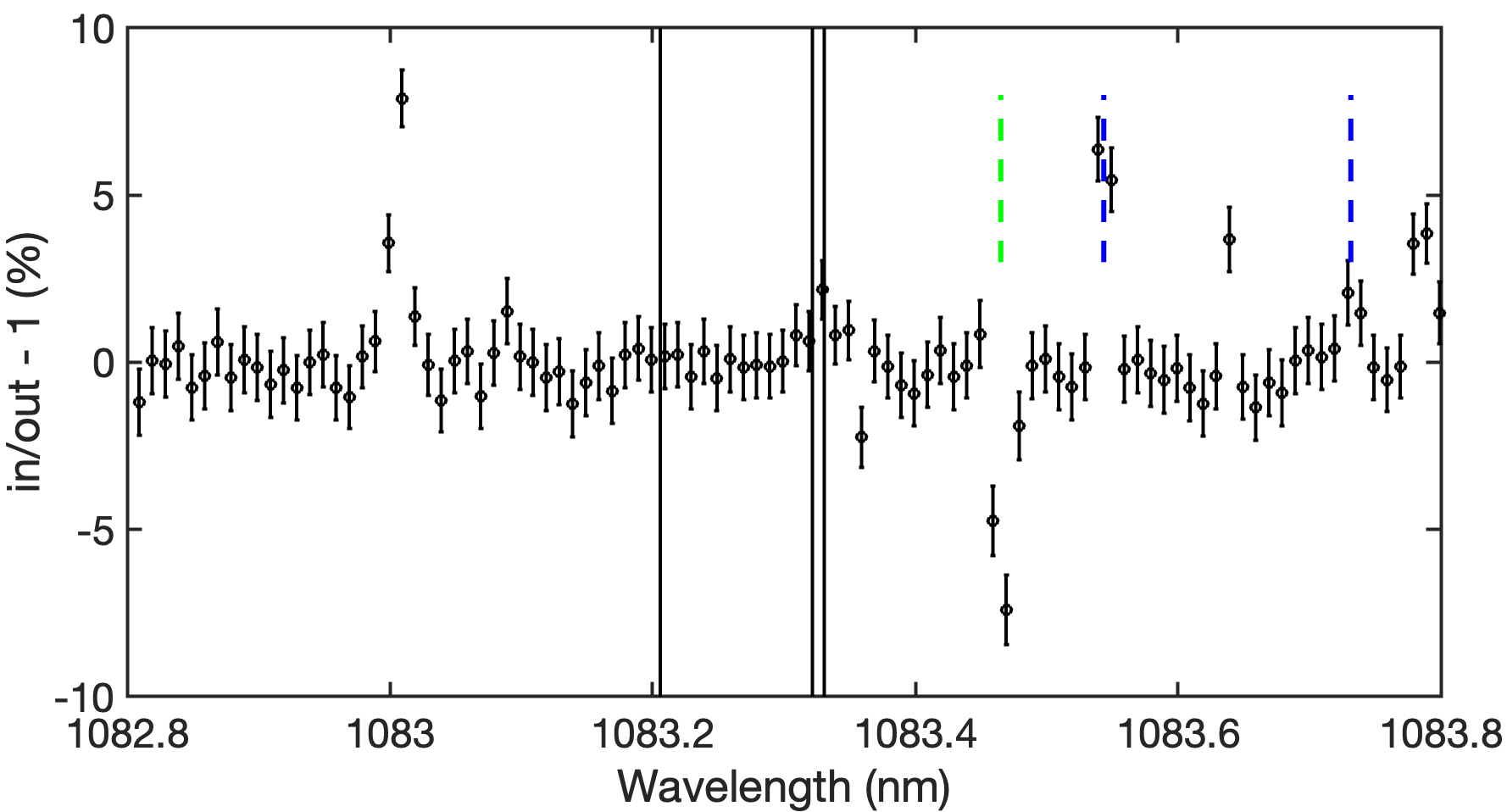}
\caption{Transmission spectrum of TOI-1235 b in planet's rest frame. No planet related absorption is visibly seen. The He\,I lines are marked in black vertical lines along with the telluric lines in green and blue.}
\label{new_toitrans}
\end{figure}

However, such corrections are not necessary if there are no telluric features in the vicinity of our spectral region of study. For TOI-1235, there were no known telluric lines in the vicinity of the He\,I lines (see Figure \ref{fig4}), so we did not perform any telluric correction. We used the updated ephemeris from ground-based LCOGT photometry \citep{toi-1235b_1} for TOI-1235 b. The planet's period, mass and radius were taken from \cite{toi-1235b_2} (see Table 1). We followed similar procedure for TOI-1235 b to construct its transmission spectrum as seen in Figure \ref{new_toitrans}.

\section{Results}

\subsection{GJ 9827 system}
\label{sec_gj9827_res}

There is no prominent absorption feature seen in both GJ 9827 b and GJ 9827 d at the He\,I wavelengths (see Figure \ref{fig_new2}). Nevertheless, to constrain any planet-related absorption, we perform an upper-limit EW estimation. A region of 2\,\AA\, centered around the He\,I lines was chosen. We modeled a single Gaussian profile with thermal broadening of 1000-30,000 K (neglecting rotational broadening). For the RMS value, we chose a region of 10\,\AA\, in the vicinity of the line (1084-1085\,nm). Running a three-parameter (position, line width and depth) chi-squared analysis, we limit any planet-associated absorption to 14.71\,m\AA\, for GJ 9827 b and 18.39\,m\AA\, for GJ 9827 d at 99\% confidence. These translate to an upper-limit of 0.21\% and 1.32\%, respectively for GJ 9827 b and d, in the possible excess absorption depths at the same 99\% confidence.

\begin{figure}
\centering
\includegraphics[width=8.65cm]{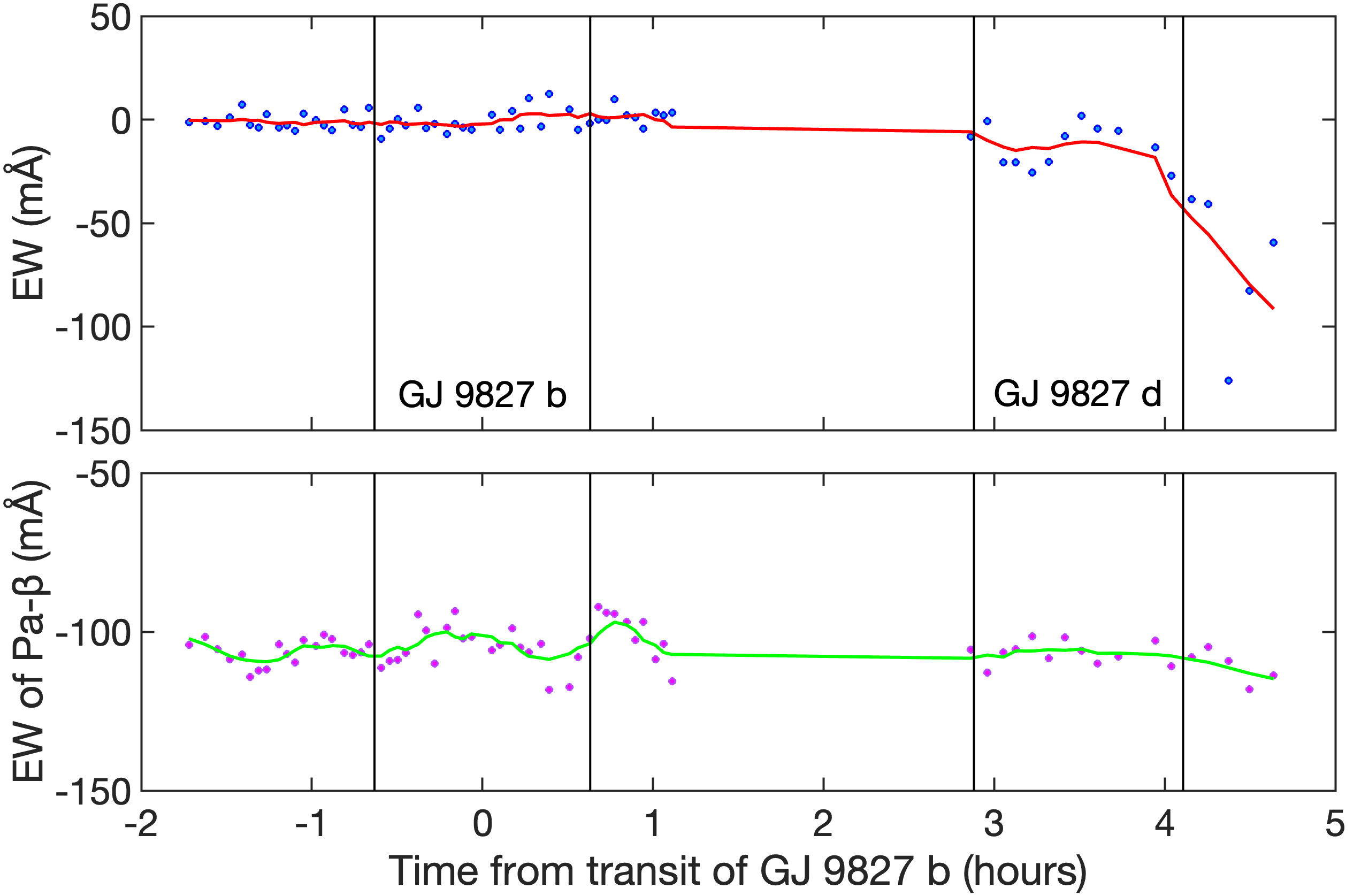}
\caption{Temporal comparison of residual flux ratio EW in the vicinity of the He\,I triplet with the Pa-$\beta$ EW. A 7-point first order Savitzky-Golay filter (red) is plotted along the raw data (blue) in the top panel. The bottom panel shows the EW of Paschen beta line seen in absorption (magenta) with the 7-point first order Savitzky-Golay filter in green. The beginning and end of GJ 9827 b and GJ 9827 d transits are marked by black vertical lines.}
\label{fig3}
\end{figure}

We chose a 2\,\AA\, region centered around the He\,I doublet and plotted its temporal variation in Figure \ref{fig3}. The mean value of the Gaussian-fitted equivalent width (EW) of this chosen 2\,\AA\, during the transit of GJ 9827 b is -0.89 $\pm$ 0.07\,m\AA\, while the out-of-transit yielded -0.76  $\pm$ 0.10 m\AA. There is no significant difference between them, indicating the absence of any excess absorption (here more negative indicates more absorption). During the transit of GJ 9827 d we get a mean EW of -12.76  $\pm$ 2.73 m\AA. This increased level is probably due to low elevation of the target, which induces instantaneous increase in the strength of telluric $\mathrm H_2$O and OH lines that coincidentally overlap the He\,I lines (see top panel of Figure \ref{fig_new1}). In such low elevations, the photons from the target pass through a larger section of high density telluric atmosphere, causing instantaneous increase in the telluric line strengths. Such a phenomenon has been previously observed \citep[e.g.,][]{kawauchi2018} from ground based telescopes. These instantaneous variations in water-vapor levels are impossible to correct for in the IRD data as it lacks a separate sky fiber that collects the simultaneous telluric measurement along with the science spectra.

The Paschen beta (Pa-$\beta$) line if seen in emission, is considered as a tracer of outflows in T-Tauri stars \citep{2004A&A...417..247W}. In our IRD spectra of GJ 9827, we see the Pa-$\beta$ line in absorption. Pa-$\beta$ line has been used as a chromospheric diagnostic tool in late type stars \citep{1998A&A...331L...5S}. This line along with H-$\alpha$ can probe various levels of chromospheric pressures. It is relatively free of telluric lines and hence can be probed simultaneously with the He\,I lines. The absorption of Pa-$\beta$ (at 1282.2\,nm) during transit can be seen as a mode of populating the $\mathrm H^+$ ions in the upper atmospheres. And in essence, if there is actual absorption, it should correlate with the He\,I lines. This Pa-$\beta$ weakly correlates with the EW of GJ 9827 (see Figure \ref{fig3}) in the vicinity of He\,I triplet lines (Pearson's R = 0.2505 and p = 0.0425; Spearman's $\rho$ = 0.1027 and p = 0.4109). Nonetheless, the significance of this weak correlation is yet to be understood \citep{our_paper_trappist1}. We also compute the seven-point first order Savitzky-Golay filter \citep{savitzkygolay} to visualize the smoothened data. This filter can smoothen the sudden abnormal variation in the data. The smoothened data also doesn't show any significant correlation.

\subsection{TOI-1235 system}

\begin{figure}
\centering
\includegraphics[width=8.65cm]{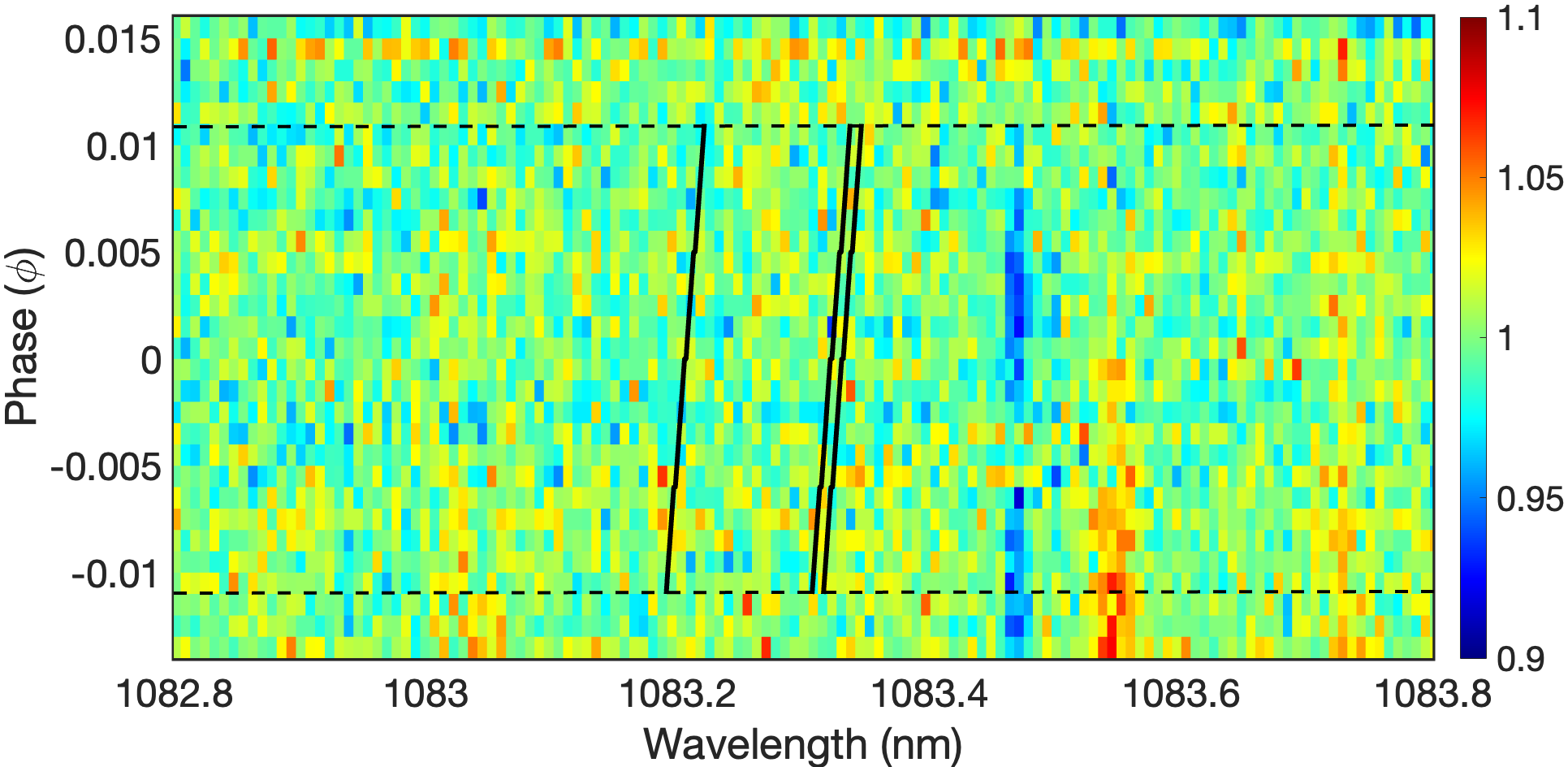}
\caption{Phase vs. wavelength map of TOI-1235 b's flux ratio (in/out) in the stellar rest frame. The black lines correspond to He\,I line positions during the transit of TOI-1235 b. The horizontal black dashed lines mark the transit contact points. No excess absorption is seen in the vicinity of He\,I lines.}
\label{toi_phase}
\end{figure}

\begin{figure}
\centering
\includegraphics[width=8.65cm]{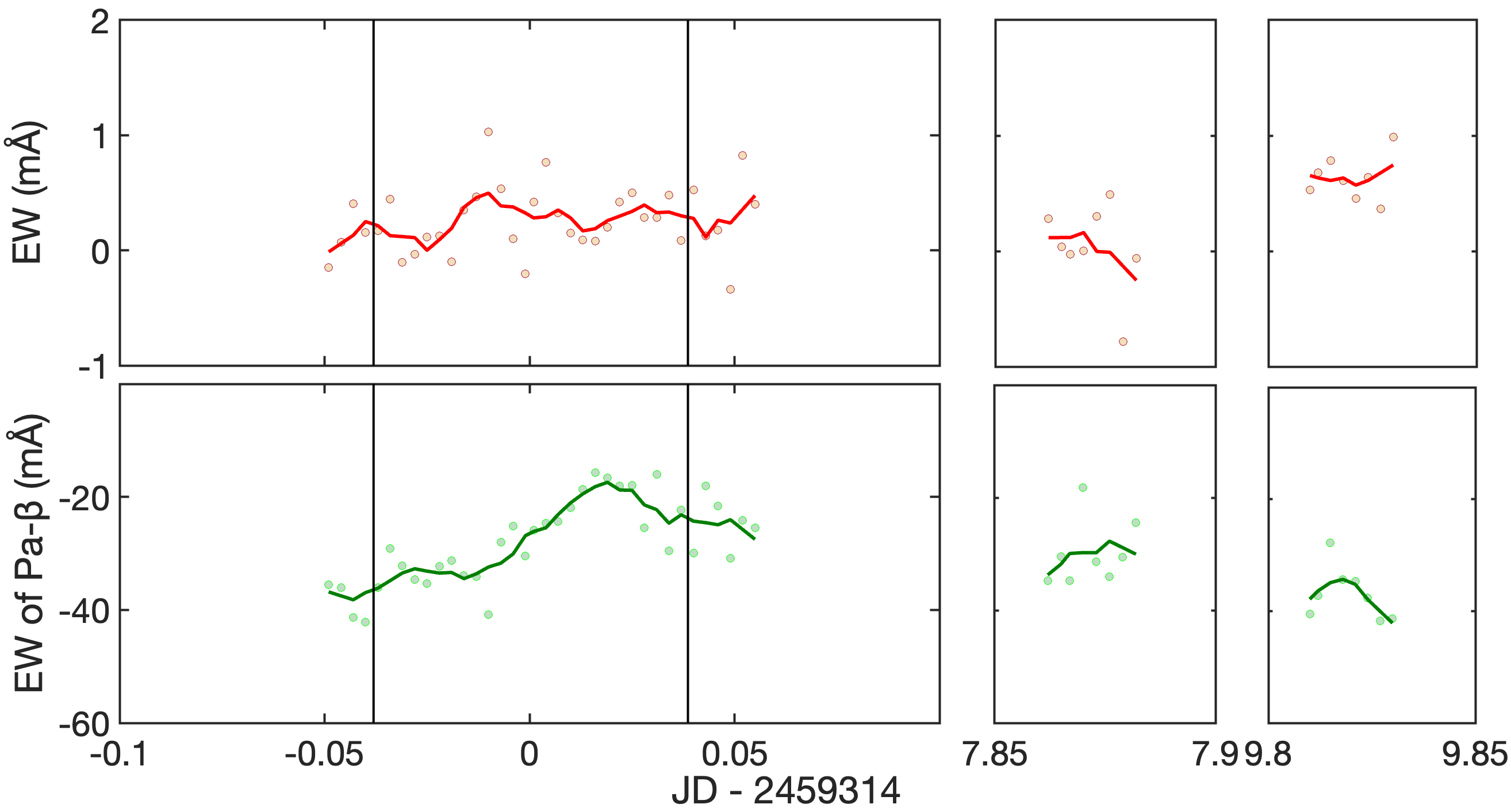}
\caption{Top: Strength of He\,I (from flux ratio) vs. time for TOI-1235. The solid red lines represent 7-point, first-order Savitzky-Golay filtered versions. Bottom: EW of stellar Hydrogen Paschen beta line (m\AA) seen in absorption. The black vertical lines mark the beginning and end of TOI-1235 b's transit on UT 2021 April 09.}
\label{fig5}
\end{figure}

The transmission spectrum of TOI-1235b is shown in Figure \ref{new_toitrans}. There are no known telluric lines present in the vicinity of the He\,I lines. So no telluric corrections were performed. Also, no excess absorption was detected from the transmission spectra. The temporal variation of the flux ratio in the vicinity of He\,I Lines as seen in Figure \ref{fig5} top panel show a large scatter before, during and after the transit. This is probably because the target was in lower elevation throughout the observation on the night of UT 2021 April 09. Also, no excess absorption is seen in the phase plot of TOI-1235 b as seen in Figure \ref{toi_phase}. By performing a similar chi-squared analysis on the flux ratio spectra as shown in Section \ref{sec_gj9827_res}, we limit any planet-related absorption in the He\,I region to be less than 1.44\,m\AA\, at 95\% confidence. In terms of excess absorption depth, this upper-limit translates to 0.09\% at the same confidence level. 

The EW of Paschen beta line seen in absorption is plotted in the bottom panels of Figure \ref{fig5}. There is no strong correlation between the He\,I triplet lines and the Pa-$\beta$ line. The raw data yields a Pearson's correlation of R = 0.0263 and a p-value of 0.8789; and Spearman's $\rho$ = 0.1308 and a p-value = 0.4457. However, there is a hint of negative correlation between He\,I and Pa-$\beta$ outside the transit, giving us a Pearson's correlation of R = -0.4886 and p-value of 0.0548; and Spearman's $\rho$ = -0.4882 and p-value of 0.0572. Comparing He\,I lines with Pa-$\beta$ during the transit yields us a weakly insignificant correlation (Pearson's R = -0.0865 and p-value = 0.6744).

\subsection{He I modeling}

Our upper-limits on the EW of any anomaly in the He I line associated with the transits of these planets were converted into limits on the rate of atmosphere escape using a 1-d model of a  symmetric isothermal Parker wind with a solar composition. The model is more fully described in \cite{Gaidos2020} and only a few important aspects are presented here.  In low-density planetary winds, the metastable triplet (2$^3$S) state -- the lower state of the He\,I transition -- is mostly populated by recombination of He ionized by extreme ultraviolet (EUV) photons with energies $>26.4$ eV  or wavelengths $\lambda < 504$\AA; loss from this state is primarily via ionization by near ultraviolet (NUV) photons with energies $>4.8$ eV \citep[$\lambda < 2583$ \AA,][]{oklopcichirata}.  To calculate the abundance of absorption by triplet He I in the wind, a spectrum of the host star over this entire UV wavelength range is needed, and since none is available for these stars we used composite spectra from analog stars selected from the MUSCLES survey \citep{muscles1, muscles2, muscles3}. We chose HD 85512, a K6V star as an analog for GJ 9827, scaled using Ly $\alpha$ and Mg II emission from \citet{gj9827_helium}. For TOI-1235 we chose a M2V star, GJ 832 scaled using \emph{GALEX} measurements \citep{galex1, galex2}, i.e. the FUV and NUV magnitudes from \citet{galex_revised}.  We adopt planet masses and radii from \cite{gj9827_plan_para} for GJ 9827 planets and from \citep{toi-1235b_1, toi-1235b_2} for TOI-1235 b.  

\begin{figure}
\centering
\includegraphics[width=8.65cm]{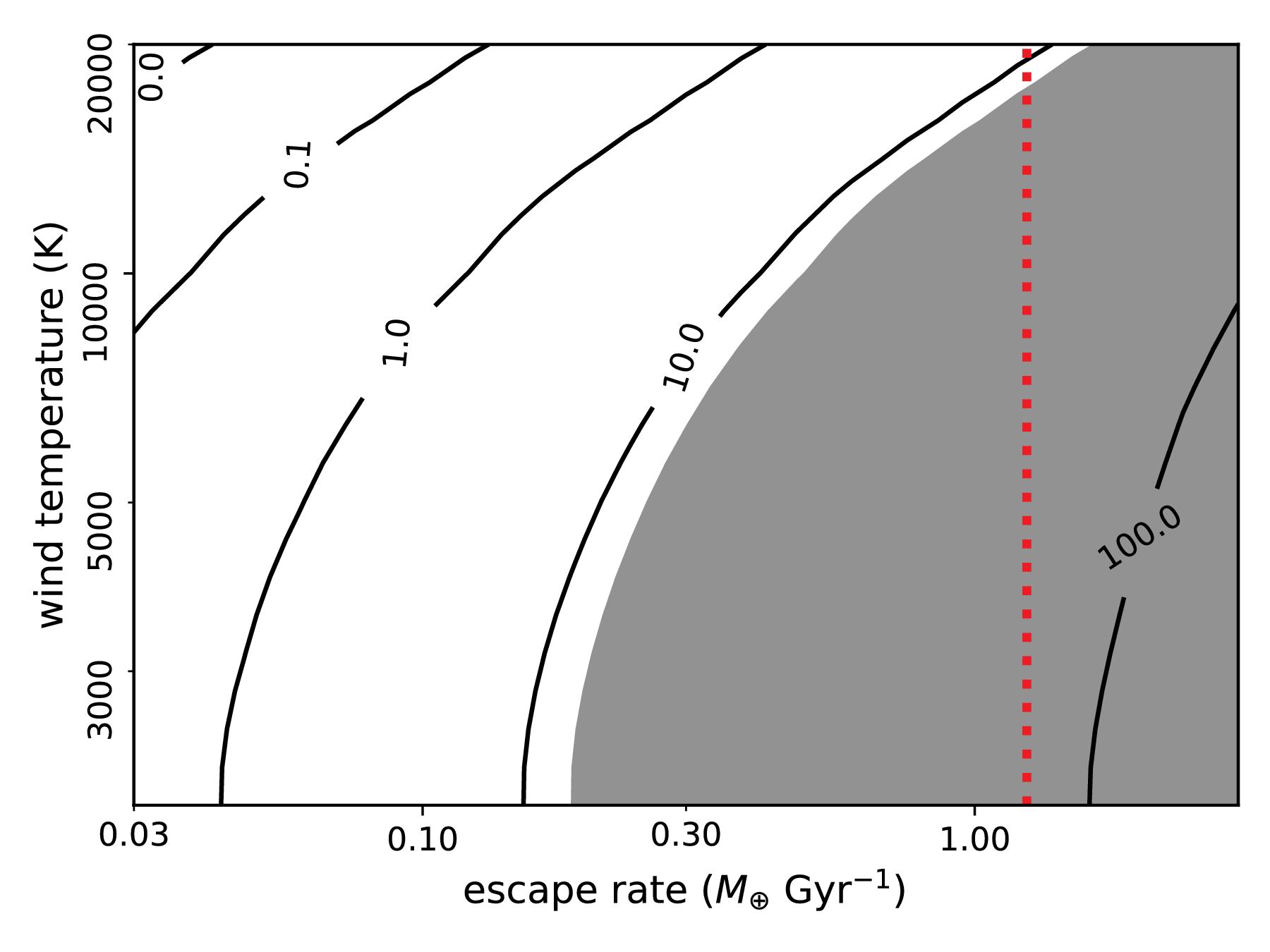}
\caption{Constraints on the atmospheric escape from GJ 9827 b. Contours of predicted EW (in m\AA) of the 1083.3\,nm triplet He\,I line versus escape rate in $\mathrm M_{\oplus}$\,Gyr$^{-1}$ and wind temperature from a solar-metallicity, isothermal Parker wind model. The grey shaded region denotes EW values $>$14.71 m\AA\ excluded by our observations at 99\% confidence. The vertical red dotted line is the theoretical prediction from \citet{kubyshkina2018}.}
\label{fig6}
\end{figure}

\begin{figure}
\centering
\includegraphics[width=8.65cm]{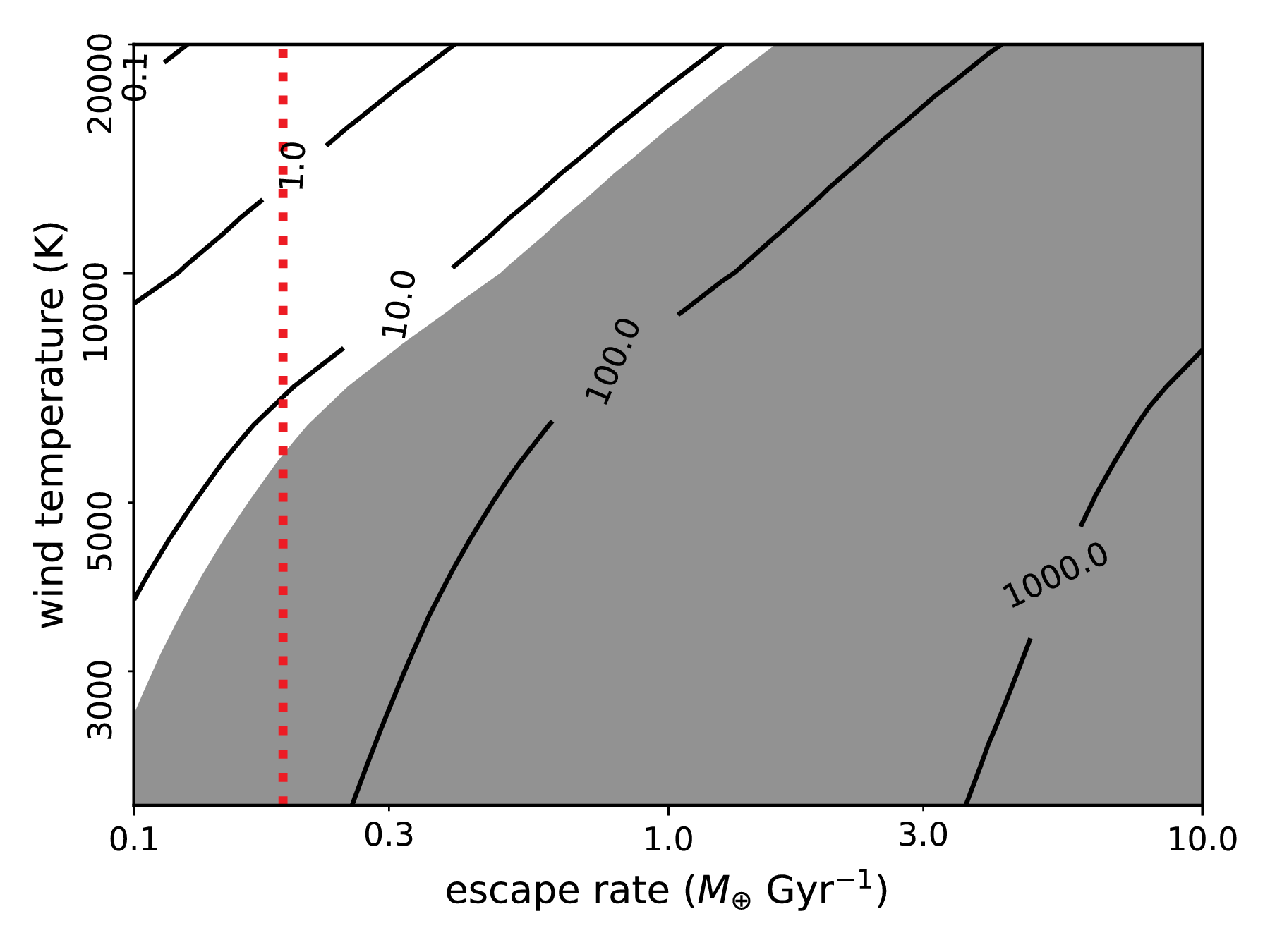}
\caption{Same as Fig. \ref{fig6} but for GJ 9827 d. The grey shaded region denotes EW values $>$18.39 m\AA\ excluded by our observations at 99\% confidence.}
\label{fig7}
\end{figure}

\begin{figure}
\centering
\includegraphics[width=8.65cm]{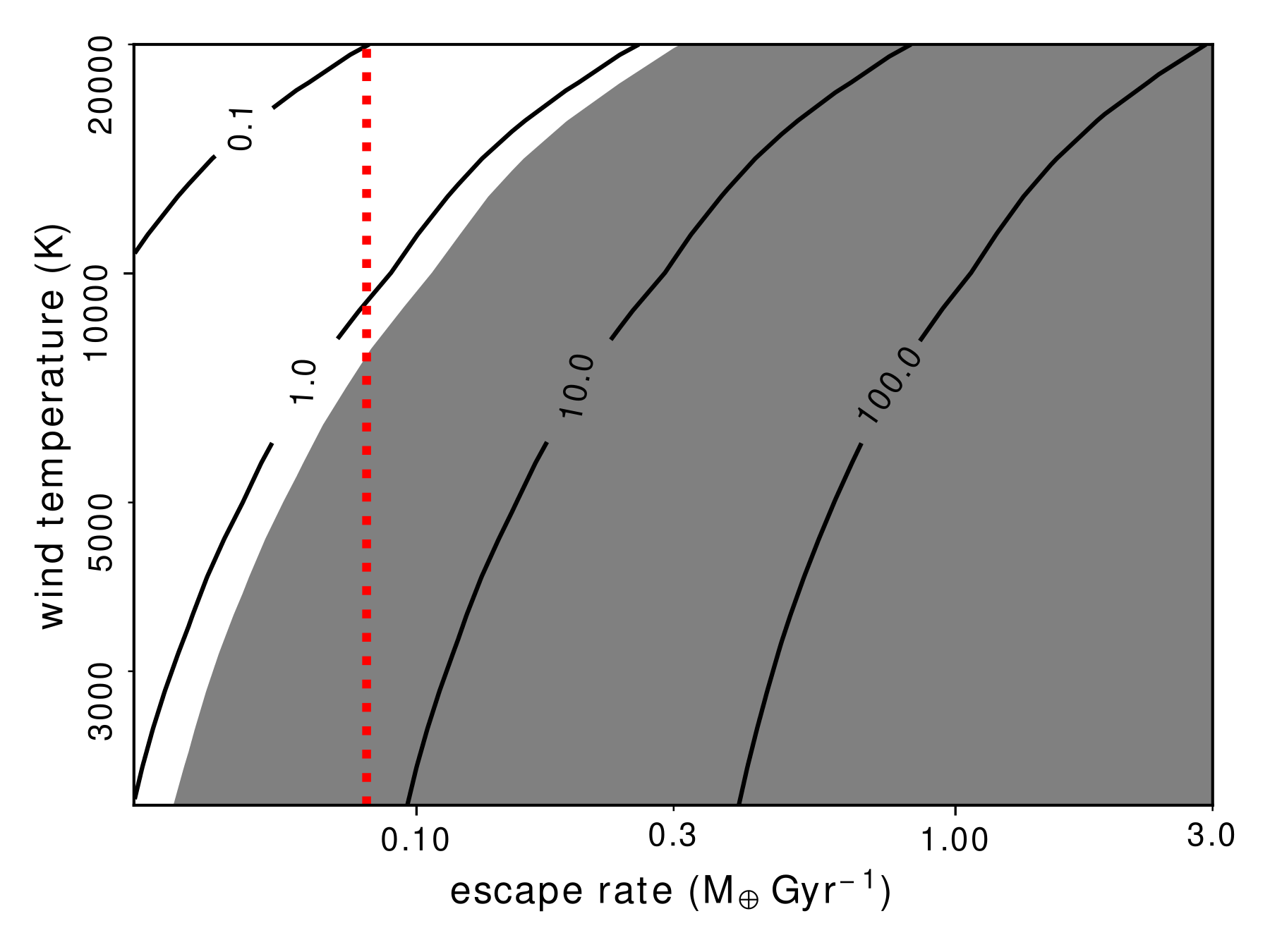}
\caption{Same as Fig. \ref{fig6} but for TOI-1235b. The grey shaded region denotes EW values $>$1.44 m\AA\ excluded by our observations at 95\% confidence.}
\label{fig8}
\end{figure}

Neutral and ionized H, singlet and triplet neutral He, ionized He, and electrons are tracked in the model. Integrated line profiles account for finite optical depth, intrinsic and thermal broadening, and the resolution of the instrument. Figures \ref{fig6}, \ref{fig7} and \ref{fig8} plot predicted EW as a function of escape rate and wind temperature.  For a representative wind temperature of 5000K, our non-detection rules out escape rates $>$0.25 $\mathrm M_{\oplus}$\,Gyr$^{-1}$ and $>$0.2 $\mathrm M_{\oplus}$\,Gyr$^{-1}$ for GJ 9827 b and d, respectively (99\% confidence), and $>$0.05 $\mathrm M_{\oplus}$\,Gyr$^{-1}$ for TOI-1235 b (95\% confidence). 

We predict energy-limited escape rates of $\sim$0.03 $\mathrm M_{\oplus}$\,Gyr$^{-1}$, $\sim$0.01 $\mathrm M_{\oplus}$\,Gyr$^{-1}$ and $\sim$0.02 $\mathrm M_{\oplus}$\,Gyr$^{-1}$ for GJ 9827 b, GJ 9827 d and TOI-1235 b respectively, using the relation in \cite{energylimited1} with a correction factor of 1 \citep[K=1;][]{energylimited2} and nominal efficiency of 10\% \citep{energylimited3}. The constraints from our limit on He~I EW and Parker wind model are significantly above these theoretical estimates.  However, escape rates based on the hydrodynamic simulations of \cite{kubyshkina2018} are more in tension with the observations:  We estimate 1.24 $\mathrm M_{\oplus}$\,Gyr$^{-1}$ and 0.19 $\mathrm M_{\oplus}$\,Gyr$^{-1}$ for GJ 9827 b and d respectively; and 0.08 $\mathrm M_{\oplus}$\,Gyr$^{-1}$ for TOI-1235 b.  These theoretical estimates are plotted as the vertical dashed red lines in Figures \ref{fig6}-\ref{fig8}.  

\section{Discussions}
\label{discussion}

\subsection{Telluric correction}

In the case of GJ 9827, we removed the telluric features by constructing a telluric template from our rapid-rotating star observations. However, since the rapid-rotator and the target (i.e., GJ 9827) were not simultaneously observed, the instantaneous correction of telluric variation becomes difficult, especially in the target is in low elevation or observed during astronomical twilight \citep{kawauchi2018}. Here, we propose an alternate method to correct the telluric feature using two-step iterative algorithm. 

\subsubsection{{\tt SYSREM} telluric correction}

The first step involves an additional observation of a rapidly rotating bright A/B standard star near the target's airmass \citep{telluric_winn2004}. We observe this rapid-rotator at the beginning or end of our science observations of the transit night to reduce the time on slewing and target acquisition \citep{telluric_narita2005, telluric_redfield2008}. These fast rotating stars broadens any absorption lines present in them \citep{1929MNRAS..89..222S}, thereby making them essentially a flat continuum spectra. Upon entering Earth's atmosphere, the telluric lines are imprinted on their spectra, turning them into an instantaneous telluric template at that particular airmass. Since the primary targets are observed at different airmasses, we transfer the telluric template to the primary target's observed airmass at each frame. This is done using Equation 1 in \cite{kawauchi2018}. We then divided each spectrum of our target by the transferred telluric template. This step essentially removes most of the telluric spectra from our target's spectra. 

\begin{figure}
\centering
\includegraphics[width=8.65cm]{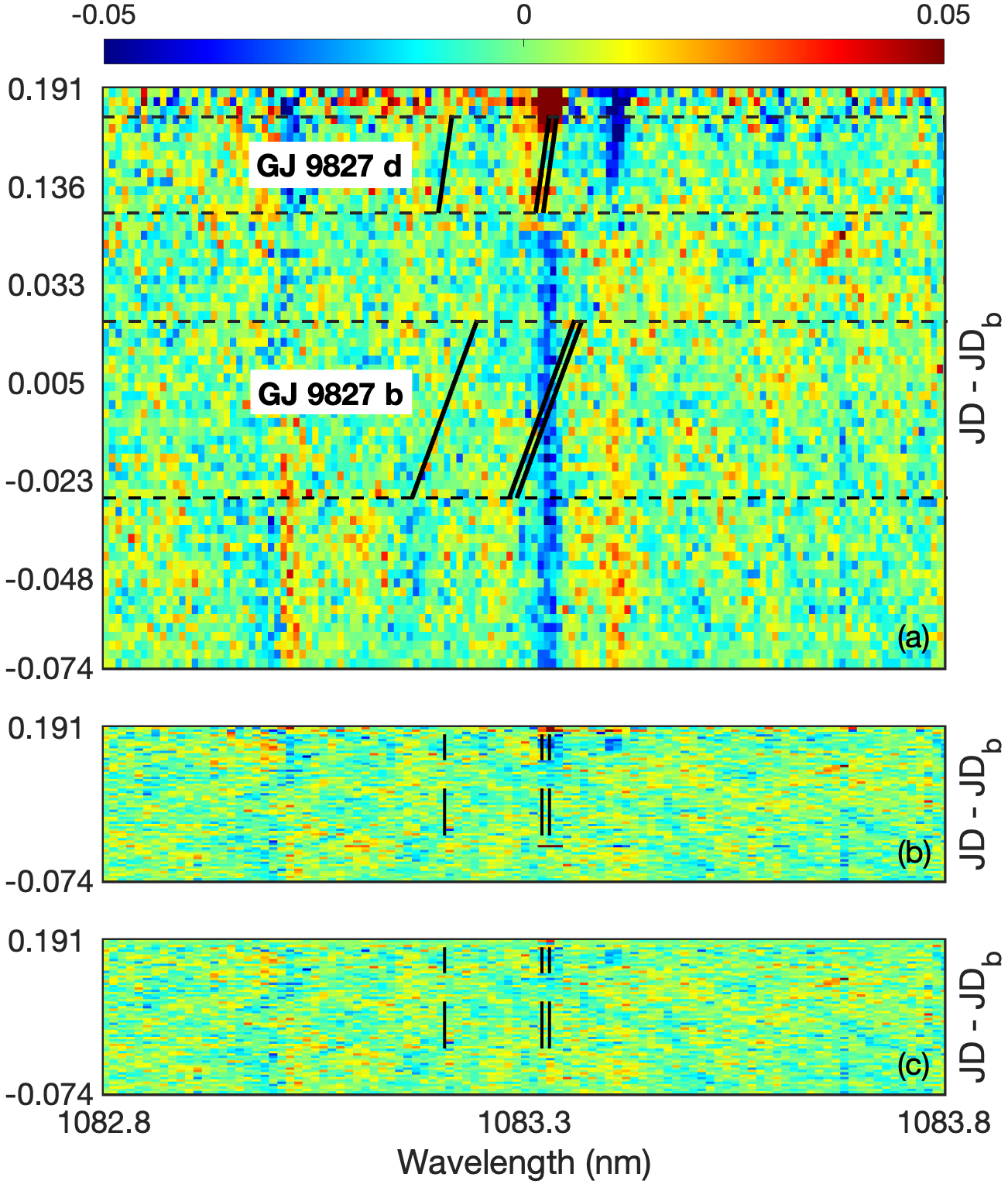}
\caption{Implementation of {\tt SYSREM} algorithm to correct for the telluric contamination in GJ 9827. The panel (a) shows the input to SYSREM algorithm (transferred to stellar rest frame for visualization). The black lines correspond to He\,I line positions during the transit of GJ 9827 b and d. The horizontal black dashed lines mark the transit contact points. The residual after airmass correction and 1$^{st}$ iteration of {\tt SYSREM} , in the planet's rest frame is shown (b). The panel (c) show the residuals after 5$^{th}$ iteration of {\tt SYSREM}. Here JD$_b$ = 2459098.932 is the Julian Date centered at the mid-transit time of GJ 9827 b.}
\label{fig1}
\end{figure}

Unfortunately, this step does not completely remove the telluric spectra. The instantaneous changes in the Earth's atmospheric wind velocity, water vapour column density, etc., induce variations that cannot be accounted with just the airmass correction method \citep{telluric_recent2021}. And in the near infrared wavelengths, the OH emission lines are highly dominant than the oxygen absorption lines, especially near the He\,I lines \citep{2018Sci...362.1388N}. Their instantaneous variations are tricky to correct by just using the rapid-rotator spectrum. Hence we proceed with a detrending algorithm {\tt SYSREM} \citep{sysrem1, sysrem2} for the removal of the residual telluric features. {\tt SYSREM} fits a systematic trend in the wavelength direction, which accounts for the various instantaneous variations mentioned above. This method have been implemented for telluric correction of high resolution exoplanetary observations \citep{2020MNRAS.493.2215G, 2021ApJ...910L...9N} by running it for the individual spectral order of interest. However, one must take caution in the number of iterations while implementing {\tt SYSREM}. Higher number of iterations might remove the planetary signals, if any, and might leave behind artifacts mimicking a planetary signal \citep{2021ApJ...910L...9N}. In the case of smaller planets like our targets, {\tt SYSREM} finds it difficult differentiate between the telluric signal and signals of planetary origin \citep{2020MNRAS.493.2215G, 2020A&A...636A.117M}. Also since there is no constrain on the number of the iterations of {\tt SYSREM} needs to be run, we performed our analysis on the residual spectra after iteration of {\tt SYSREM}. Also, {\tt SYSREM} might not completely remove the telluric features if the planet-induced RV does not spread much in the wavelength direction \citep{2020MNRAS.493.2215G}. This {\tt SYSREM}-based two-step telluric correction for GJ 9827 is shown in Figure \ref{fig1}. 

We followed the steps from \cite{2020MNRAS.493.2215G} and \cite{2021ApJ...910L...9N} and performed the iterations while correcting the telluric features in real GJ 9827 spectra. The number of iterations in {\tt SYSREM} can strongly determine the nature of residual. Hence in order not to lose any useful planetary signals, if any, we performed our analysis after each iteration of {\tt SYSREM} in GJ 9827 system. The Gaussian fit model clings to the residual after the first iteration, when compared to the successive iterations. {\tt SYSREM} can perform better for large, heavier planets in the close-in orbits as the algorithm looks for temporal changes in the spectra. GJ 9827 planets, being in not so close-in orbits, might not show significant temporal changes in the observer's rest frame \citep{2020MNRAS.493.2215G}.  Hence the planetary signals if any, might have been removed in the telluric corrections. Nevertheless, on looking at the spectra just after the airmass telluric correction and before the {\tt SYSREM} iterations, one can clearly see the strong residuals of the telluric features, hindering in our analysis. Hence this second step telluric correction is an essential tool in completely removing the telluric features. However, {\tt SYSREM} cannot account for very large telluric features (e.g., higher concentration of OH during astronomical twilight), when compared to the continuum level. \cite{2021ApJ...910L...9N} suggests masking the strong telluric lines before the analysis. But in our observation of GJ 9827, the telluric OH line falls right in the middle of the He\,I line, rendering us from masking.

\begin{figure*}
\centering
\includegraphics[width=17.8cm]{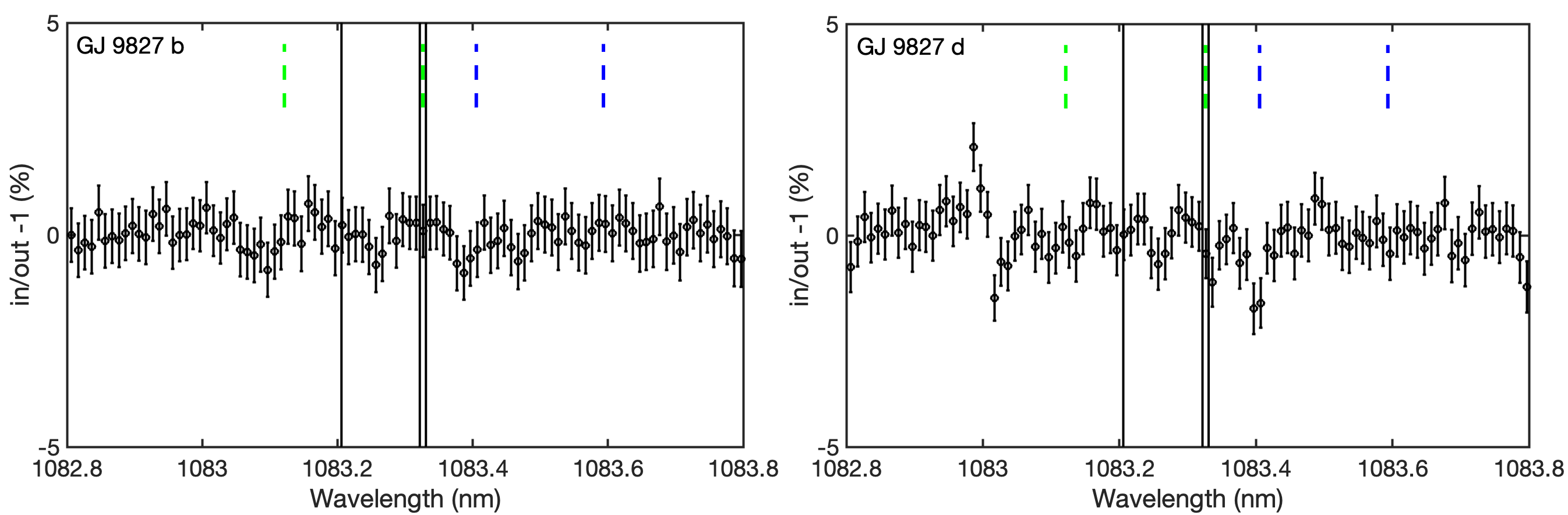}
\caption{Transmission spectra of GJ 9827 planets from the flux residuals after the first iteration of SYSREM in the planet's rest frame. The telluric OH emission lines shown in green \citep[computed from][]{telluric_oh} shows a strong line falling right in the middle of the He\,I doublet. The computed known telluric $\mathrm H_2$O (blue) positions \citep[from][]{telluric_h2o} do not fall in the vicinity of He\,I lines. We do not see any significant absorption in both the planets.}
\label{fig2}
\end{figure*}

The transmission spectra near the He\,I lines generated from the residuals after the telluric corrections for GJ 9827 b and GJ 9827 d are shown in Figure \ref{fig2}. The rms of the residuals after each iteration reduces drastically, thereby possibly removing the planetary signals, if any present \citep[e.g.,][]{zhang2021}. However with just after the airmass correction using the rapidly rotating star, the telluric features are still very prominent. Therefore, further analyses are performed after incorporating the {\tt SYSREM} algorithm. 

We performed similar statistics and estimated upper-limits on the EWs of any planet-related absorption. For GJ 9827 b, we place an upper-limit of 13.12\,m\AA\, at 99\% confidence and for GJ 9827 d we constrain the EW to 16.38\,m\AA\, at 99\% confidence. In terms of excess absorption depth, these correspond to 0.15\% and 0.79\% respectively. These limits are slightly lower than the estimated limits from the standard telluric correction method described in Section \ref{new_tell_sec}. This could indicate the additional removal of telluric features by {\tt SYSREM}. These upper-limits translate to similar mass-loss limits using our models. This telluric correction was not performed for TOI-1235 spectra as there were no telluric lines in the vicinity of He\,I lines (see Figure \ref{fig4}).

\begin{figure}
\centering
\includegraphics[width=8.65cm]{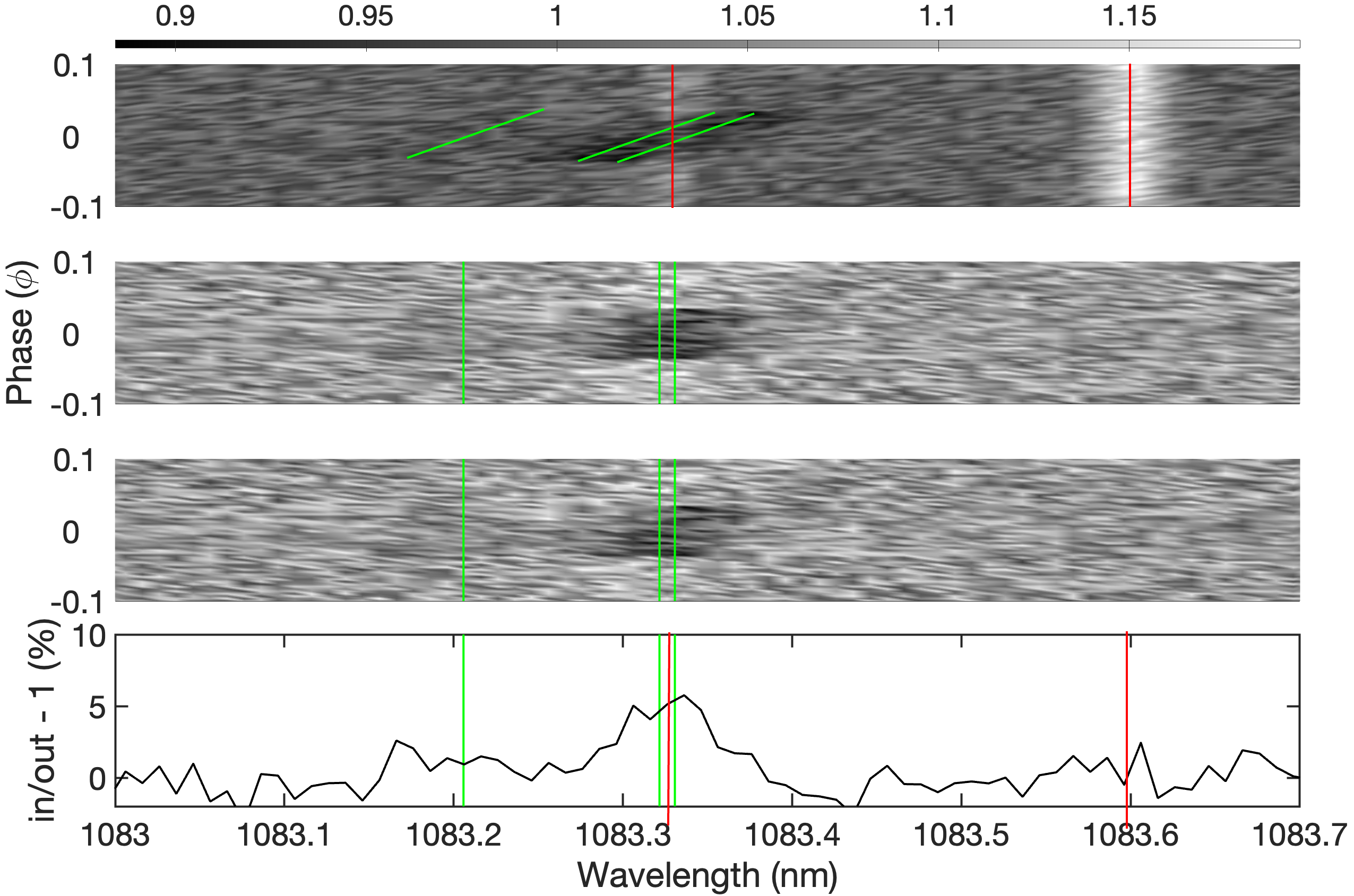}
\caption{Telluric removal on synthetic transit spectra of a planet similar to that of GJ 9827 b. The top panel shows the raw data inputted to {\tt SYSREM} in the stellar's rest frame. Synthetic telluric OH emission lines (red lines) are added at the center of He\,I doublet line and away from the He\,I lines at 1083.6\,nm. The second and third panel shows the residuals after one and three iterations of {\tt SYSREM} respectively in the planet's rest frame. The bottom panel shows the 1-d collapsed flux ratio of the resides after the third iteration. }
\label{fig11}
\end{figure}

To demonstrate the validity of {\tt SYSREM}, we injected a 10\% absorption He\,I signal for a mock-planet like GJ 9827 b. We performed a similar analysis, and were able to retrieve back the signal after three iterations of {\tt SYSREM} (see Figure \ref{fig11}). The telluric features were completely removed in both the locations and we were able to retrieve back an absorption signal of $\approx$6.1\%. The lost He\,I could have been over-corrected by the {\tt SYSREM} algorithm, mistaking it for telluric signal as the planet does not undergo extreme radial acceleration during the observation \citep{zhang2021}. This method is particularly ideal for spectrographs like IRD on Subaru telescope, which does not have a separate sky-fiber like in ESPRESSO on the VLT \citep{2014AN....335....8P}. 

\subsection{Helium analysis}

\begin{figure}
\centering
\includegraphics[width=8.65cm]{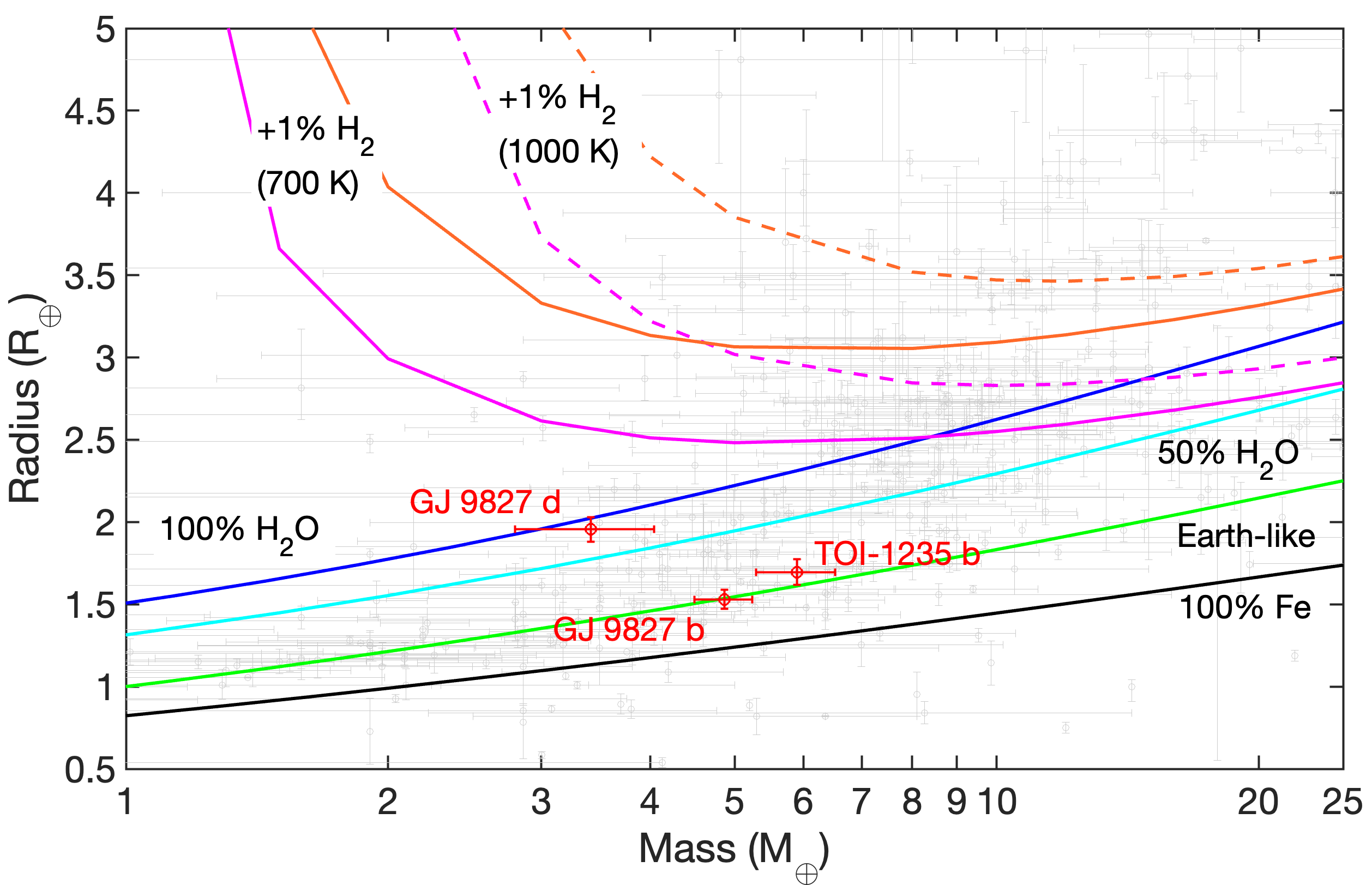}
\caption{Mass-radius plot for known exoplanets \citep{2011A&A...532A..79S}. The contours are for various composition models like pure iron (black; 100\% Fe), Earth-like (green; 32.5\% Fe + 67.5\% silicates), 50\% water world (cyan; 50\% Earth-like + 50\% by mass $\mathrm H_2$O), 100\% water world (blue), 1\% $\mathrm H_2$ on Earth-like (solid magenta - 700\,K; dashed magenta - 1000\,K) and 1\% $\mathrm H_2$ on Earth-like + water (solid orange - 700\,K; dashed orange - 1000\,K) from \citet{2019PNAS..116.9723Z}. GJ 9827 b might have a barren rocky core, while GJ 9827 d and TOI-1235 b might host some volatiles.}
\label{fig9}
\end{figure}

The neutral Helium is not detected in the atmospheres of GJ 9827 b, GJ 9827 d, and TOI-1235 b (see Figures \ref{fig_new2} and \ref{new_toitrans}). The GJ 9827 planets lie on either side of the radius-gap, with planet `b' expected being rocky and planet `d' expected to have an atmosphere that is more prone to escape \citep{gj9827_plan_para}. The isochrone analysis predicts the system to be old \citep{gj9827_mass1}, indicating the hydrodynamic atmospheric escape of primordial atmosphere might have ended in the first few hundred million years \citep{OwenWu2013, OwenWu2017}. This could have paved way for the formation of secondary atmosphere (perhaps $\mathrm H_2$O dominated). And since this was during the most active time of the early star, the secondary atmospheres could have also been evaporated, leaving behind a barren rock \citep{2011ApJ...742L..19M}. This could be the case of GJ 9827 b as its mass and density indicate that it might have an Earth-like internal composition (see Figure \ref{fig9}). And the mass and radius of GJ 9827 d indicate that it might be a world with dominant water atmosphere. The Figure \ref{fig9} is constructed using the current mass-radius exoplanet data from the exoplanet.eu database \citep{2011A&A...532A..79S}. Different planet composition models are taken from \cite{2019PNAS..116.9723Z} at 500\,K. The various models presented are pure iron (black; 100\% Fe), Earth-like (green; 32.5\% Fe + 67.5\% silicates), 50\% water world (cyan; 50\% Earth-like + 50\% by mass $\mathrm H_2$O), 100\% water world (blue), 1\% $\mathrm H_2$ on Earth-like (solid magenta - 700\,K; dashed magenta - 1000\,K) and 1\% $\mathrm H_2$ on Earth-like + water (solid orange - 700\,K; dashed orange - 1000\,K). The lower EW of He\,I seen after the egress of GJ 9827 d could be from the low-elevation of the target and not indicating a planetary tail \citep[e.g.,][]{2018Sci...362.1388N}. Recent observations \citep{kasper2020gj9827d, gj9827_helium} also show no excess absorption, confirming our results.

The age of TOI-1235 is not accurately determined, however \cite{toi-1235b_2} gives a broad range of 0.6--10 Gyr. And since the photoevaporation of primordial atmosphere only takes a few hundred million years, this non-detection of He\,I is not a surprise even for the lower age limit. The planet might have developed thinner secondary atmosphere after the stripping of its primordial atmosphere, indicative of its density as seen in Figure \ref{fig9}. 

The Pa-$\beta$ line is seen in absorption for both targets. However, during the observation of GJ 9827, it was temporally invariant, indicating that there is no significant contribution from surface features (i.e., the emission is relatively uniform over the stellar disk). Its marginal correlation with the He\,I can be from the large scatter in the data and observational noise. The increase during the transit of TOI-1235 b might be seen as a partial indication of excited Hydrogen atoms, however it could not be clearly established from the He\,I comparison. Understanding the physical significance of this Pa-$\beta$ line could be of utmost importance, as it provides a direct tool to explore the atmospheric escape along with He\,I lines from the same wavelength regime.

Our Parker wind model assumes a wind of solar composition, which might not be appropriate for small planets with evolved atmospheres that have lost H/He or accreted with metal-rich atmospheres.  Also we substituted the EUV fluxes of these stars with the scaled fluxes from the MUSCLES database \citep{muscles1, muscles2}. In the case of GJ 9827, we scaled the  HD 85512's spectra with the only measured spectra, i.e., Ly-$\alpha$ (from HST) and Mg\,II emission lines (from CARMENES) from \cite{gj9827_helium}. We used the flux density of the reconstructed intrinsic stellar Ly $\alpha$ line at 1216 \AA. The theoretical mass-loss model derived from \cite{kubyshkina2018} for GJ 9827 b is on the higher end of 1.24 $\mathrm M_{\oplus}$\,Gyr$^{-1}$. Being in an old system, this planet might have lost its primordial H/He atmosphere in the initial few hundred million years \citep{OwenWu2013}. The star might have been still active when it lost its then formed secondary atmosphere. This high mass-loss rate could originate from a metal-rich exosphere \citep{2011ApJ...742L..19M}. The low density of GJ 9827 d \citep[2.41 g cm$^{-3}$;][]{gj9827_plan_para} and the lower predicted theoretical escape rates indicate that the planet might still host volatiles in its atmosphere. However, the lack of detection from He\,I lines indicate that the atmosphere is not H/He dominated. The planet might have incorporated molecular volatiles like $\mathrm H_2$O, C$\mathrm O_2$ etc. above its metal-rich core, bringing the density down to the measured values. The predicted values from our paper matches with the recent CARMENES measurements of He\,I \citep{gj9827_helium}. We predicted the time of transit for GJ 9827 planets on the observation night from the propagated ephemeris in NASA Exoplanet Archive using the transit ephemeris from \cite{gj9827_mass1}. This shows large uncertainty in the time of transit and the results might be different if there is any drastic update in ephemeris.  

\begin{figure}
\centering
\includegraphics[width=8.65cm]{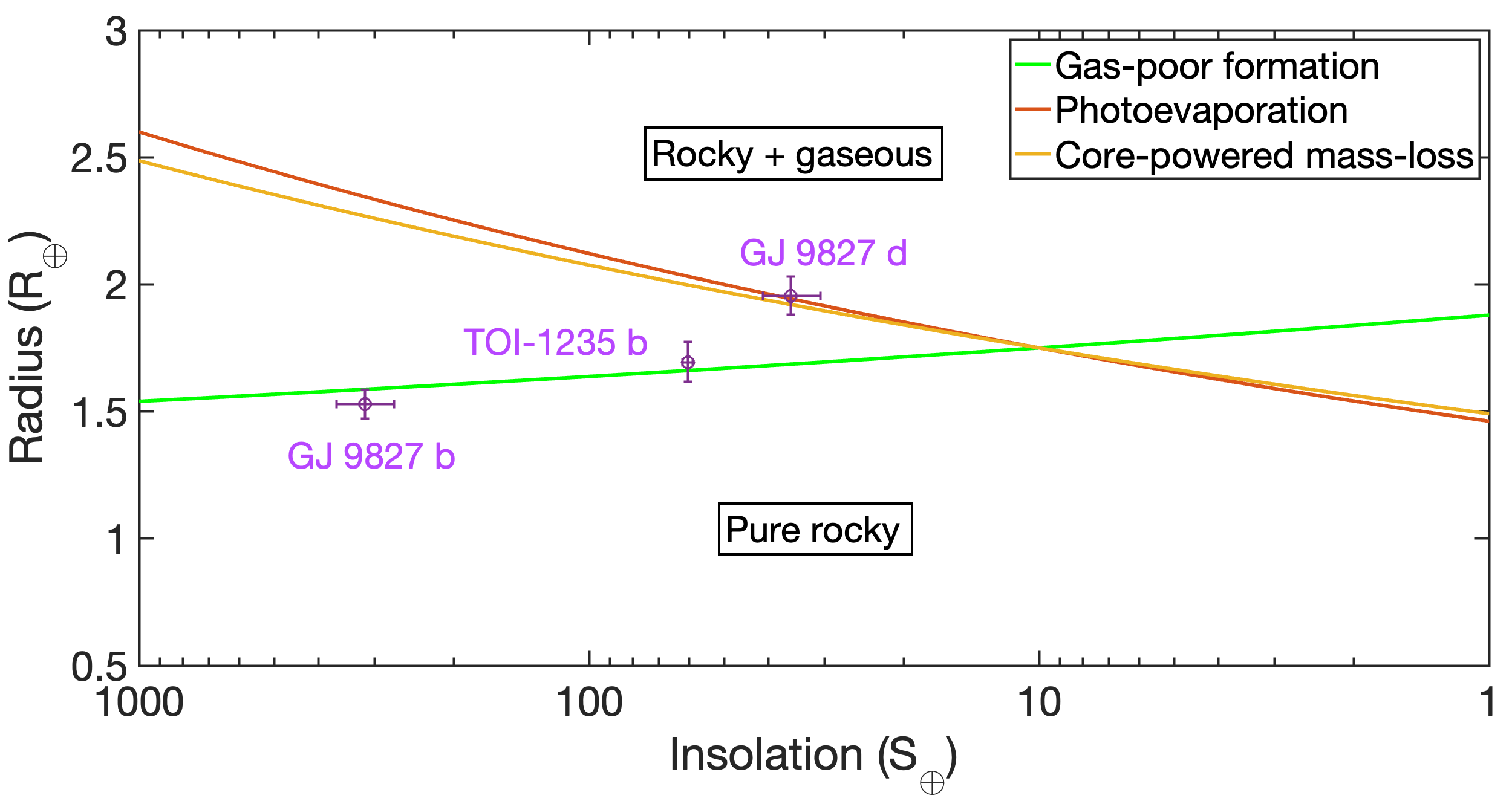}
\caption{Predictive transition from rocky to non-rocky planets in low-mass stars. The photoevaporation model \citep[red;][]{2018MNRAS.479.5303L} and core-powered mass-loss model \citep[orange;][]{2019MNRAS.487...24G} show a positive slope with insolation. The gas-poor formation model \citep[green;][]{cloutiermenou2020} shows a negative slope with the insolation. GJ 9827 b is predicted to be rocky by all three models, while GJ 9827 d is predicted to be marginally rocky by photoevaporation and core-powered mass-loss model. TOI-1235 b has opposing predictions from gas-poor formation model and photoevaporation/core-powered mass-loss models.} 
\label{fig10}
\end{figure}

TOI-1235 b on the other hand has a density comparable to that of Earth \citep[6.7 g cm$^{-3}$;][]{toi-1235b_2}. The planet structure models revealed that TOI-1235 b has an iron mass-fraction comparable to that of Earth \citep{toi-1235b_1}. Our non-detection of exosphere Helium supports this scenario of less than 0.5\% by mass of H/He dominant atmosphere. TOI-1235 b is a keystone planet that is in the crosswinds of model predictions. Three different models predict unique transitions from rocky to non-rocky planets (i.e. the location of radius-gap) in low-mass star planets. The photoevaporation models predict the slope of radius-gap varies with insolation as $\mathrm r_{p,gap} \propto S_{\oplus}^{0.11}$ \citep{2018MNRAS.479.5303L} while the core-powered mass-loss models predict the slope to be $\mathrm r_{p,gap} \propto S_{\oplus}^{0.10}$ \citep{2019MNRAS.487...24G}. The second class of models, i.e., the gas-poor formation models predict the slope of radius-gap varies with insolation as $\mathrm r_{p,gap} \propto S_{\oplus}^{-0.06}$ \citep{cloutiermenou2020}. The predicted boundaries from these models are shown in Figure \ref{fig10}. 

The insolation values for TOI-1235 b were taken from \cite{toi-1235b_2} and for GJ 9827 planets were taken from \cite{gj9827_disc2}. The gas-poor formation model predict that TOI-1235 b might be a rocky core with a gaseous envelope. But the photoevaporation model/core-powered mass-loss model show the planet is barren rock. From our non-detection and density measurement, we can see that TOI-1235 b supports atmospheric evolution through photoevaporation and core-powered mass-loss mechanisms in low-mass star planets. All three models support the rockiness of GJ 9827 b. In the case of GJ 9827 d, the planet is exactly on the boundary predictions from photoevaporation and/or core-powered mass-loss models. A further look at the molecular bands in infrared wavelengths through JWST can support clear the discrepancies among the models. 

Mention must be made here regarding the excess flux seen at the Silicon line at 1083\,nm during the transit of TOI-1235 b (see Figure \ref{new_toitrans}) and GJ 9837 d (see Figure \ref{fig_new2}). We initially predicted it to be an artifact arising from the misalignment of Silicon line while constructing the transmission spectrum. However, on further investigation, we find the line to be aligned. This could also be due to a change in line profile arising from an astrophysical effect (e.g., star spots) or a systematic from the IRD instrument. However, both these stars are slow rotators \citep{gj9827_mass1, toi-1235b_2} which renders them from generating spots. Further analysis is essential to understand this effect, which is beyond the scope of this current study.

\section{Conclusion}

We presented the NIR transit observations of GJ 9827 b, GJ 9827 d and TOI-1235 b from the high-dispersion near-infrared IRD spectrograph on the 8.2\,m Subaru telescope. Excess absorption at 1083.3\,nm He\,I is not detected in any of the planets. We placed an upper-limit EW of 14.71\,m\AA\, for GJ 9827 b (0.21\% excess absorption depth at 99\% confidence), 18.39\,m\AA\, for GJ 9827 d (1.32\% excess absorption depth at 99\% confidence) and 1.44\,m\AA\, for TOI-1235 b (0.09\% excess absorption depth at 95\% confidence). We must be cautious in these results as the telluric absorption and emission lines fall in the He\,I triplet for GJ 9827. We also implemented a novel two-step telluric correction technique using {\tt SYSREM} for the GJ 9827 spectra to adequately remove the telluric features and compared it with the standard telluric correction methods. By this method, we also limit the EW to 13.12\,m\AA\, (0.15\% excess absorption depth) and 16.38\,m\AA\, (0.79\% excess absorption depth) for GJ 9827 b and d respectively. Similar analysis was not performed for TOI-1235 b as the He\,I lines were free from telluric contamination. We note that, one should be really careful in implementing {\tt SYSREM} telluric correction. If the planets do not show high radial acceleration during observation, the {\tt SYSREM} algorithm can essentially remove planetary signals as it cannot disentangle the telluric signals from possible planetary signals. Authors strongly insist on reader's discretion in implementing this method. The Pa-$\beta$ line can act as a tracer of chromospheric activities in the star, however, further study with more samples are needed to establish a relationship between Pa-$\beta$, He\,I and stellar activities. Our Parker wind mass-loss model caps the mass-loss to $>$0.25 $\mathrm M_{\oplus}$\,Gyr$^{-1}$ and $>$0.2 $\mathrm M_{\oplus}$\,Gyr$^{-1}$ for GJ 9827 b and d, respectively (99\% confidence), and $>$0.05 $\mathrm M_{\oplus}$\,Gyr$^{-1}$ for TOI-1235 b (95\% confidence) for a representative wind temperature of 5000\,K. GJ 9827 d has a low density comparable to $\pi$ Men c. Studies of such low density super-Earths are essential in understanding the demographics of super-Earths. Also, more samples are needed in the radius-gap regime (1.5--2 $\mathrm R_{\oplus}$) around low-mass stars to concretely predict the dominant atmospheric evolution model in low-mass star planets.
 
\section*{Acknowledgements}
VK would like to thank MEXT scholarship for supporting his graduate studies. EG was supported by NASA Grant 80NSSC20K0957 (Exoplanets Research Program). R.K acknowledges support from the GSFC Sellers Exoplanet Environments Collaboration (SEEC), which is supported by NASA's Planetary Science Divisions Research Program. This work is partly supported by JSPS KAKENHI Grant Numbers JP19K14783 and JP21H00035. M.T. is supported by JSPS KAKENHI grant Nos.18H05442, 15H02063, and 22000005. The material is based upon work supported by NASA under award number 80GSFC21M0002.

\section*{Data Availability}
The data underlying this article will be shared on reasonable request to the corresponding author.



\bibliographystyle{mnras}
\bibliography{paper_biblio} 



\appendix


\bsp	
\label{lastpage}
\end{document}